\documentclass[12pt]{article}

\pdfoutput=1

\usepackage{graphics}
\usepackage{epsfig}
\usepackage{amsfonts}
\usepackage{amssymb}
\usepackage{amsmath}
\def\a{{\alpha}}
\def\b{{\beta}}

\def\g{{\gamma}}

\def\d{{\delta}}

\def\bfone{\relax{\rm 1\kern-.35em 1}}

\newcommand{\be}{\begin{equation}}
\newcommand{\ee}{\end{equation}}
\newcommand{\ben}{\begin{displaymath}}
\newcommand{\een}{\end{displaymath}}
\newcommand{\bea}{\begin{eqnarray}}
\newcommand{\eea}{\end{eqnarray}}

\newcommand{\bean}{\begin{eqnarray*}}
\newcommand{\eean}{\end{eqnarray*}}

\DeclareMathAlphabet{\mathpzc}{OT1}{pzc}{m}{it}

\begin{document}
\pagestyle{plain}

%----------------------------------------------------------------------%
%  numbering sections, equations, footnotes, etc...
%----------------------------------------------------------------------%

\makeatletter \@addtoreset{equation}{section} \makeatother
\renewcommand{\thesection}{\arabic{section}}
\renewcommand{\theequation}{\thesection.\arabic{equation}}
\renewcommand{\thefootnote}{\arabic{footnote}}

%----------------------------------------------------------------------%
%  Resetting of counters
%----------------------------------------------------------------------%

\setcounter{page}{1} \setcounter{footnote}{0}

%----------------------------------------------------------------------%
%  title page
%----------------------------------------------------------------------%

\begin{titlepage}

\begin{flushright}
UUITP-10/15\\
IPhT-t15/086
\end{flushright}

\bigskip

\begin{center}

\vskip 0cm

{\LARGE \bf Universal dS vacua in STU-models} \\[6mm]

\vskip 0.5cm

{\bf J.~Bl{\aa}b\"ack$^1$,  U.~H.~Danielsson$^2$, G.~Dibitetto$^2$  \,and\, S.~C.~Vargas$^2$}\let\thefootnote\relax\footnote{{\tt johan.blaback@cea.fr, ulf.danielsson@physics.uu.se, giuseppe.dibitetto@physics.uu.se, sergio.vargas@physics.uu.se}}\\

\vskip 25pt

{\em $^1$Institut de Physique Th\'eorique, Universit\'e Paris Saclay, CEA, CNRS, \\ F-91191 Gif-sur-Yvette Cedex, France \\[2mm]
$^2$Institutionen f\"or fysik och astronomi, University of Uppsala, \\ Box 803, SE-751 08 Uppsala, Sweden \\}

\vskip 0.8cm

\end{center}

\vskip 1cm

\begin{center}

{\bf ABSTRACT}\\[3ex]

\begin{minipage}{13cm}
\small

Stable de Sitter solutions in minimal F-term supergravity are known to lie close to Minkowski critical points. We consider a class of STU-models arising from type IIB compactifications with
generalised fluxes. There, we apply an analytical method for solving the equations of motion for the moduli fields based on the idea of treating derivatives of the superpotential of different orders 
up to third as independent objects. 
In particular, supersymmetric and no-scale Minkowski solutions are singled out by physical reasons. Focusing on the study of dS vacua close to supersymmetric Minkowski points, we are able to elaborate
a complete analytical treatment of the mass matrix based on the sGoldstino bound. This leads to a class of interesting universal dS vacua. 
We finally explore a similar possibility around no-scale Minkowski points and discuss some examples. 

\end{minipage}

\end{center}

\vfill

\end{titlepage}

%%%%%%%%%%%%%%%%%%%%%%%%%%%%%%%%%%%%%%%%%%%%%%%%%%%%%%%%%
%%
%%               Contents
%%
%%%%%%%%%%%%%%%%%%%%%%%%%%%%%%%%%%%%%%%%%%%%%%%%%%%%%%%%%

\tableofcontents

\section{Introduction}
\label{sec:introduction}

It remains a great challenge to construct meta-stable de Sitter (dS) solutions in string theory. A useful framework is provided by $\mathcal{N}=1$ supergravity specified through the choice of a real K{\"a}hler potential $K$ and a holomorphic superpotential $W$. This is phenomenologically motivated, sufficiently restricted to provide non-trivial constraints, and general enough to allow for interesting solutions. 
To obtain a dS solution, supersymmetry needs to be broken, either through explicit breaking, \emph{e.g.} by adding an anti-brane, or through spontaneous supersymmetry breaking when F-terms are turned on.

Many of the most studied models in the literature, including those inspired by string theory, include exponential contributions to the superpotential motivated by non-perturbative effects \cite{Kachru:2003aw}. These terms play a crucial role through their contribution to the stability of the models. This is the case in \cite{Kallosh:2014oja, Marsh:2014nla}, where attempts were made to find stable dS in a constructive way through spontaneous breaking of supersymmetry. The starting point for the constructions is either no-scale or supersymmetric Minkowski (Mkw).

In no-scale theories supersymmetry is partially broken through an F-term, and the vacuum energy guaranteed to be zero thanks to a translational invariance of the superpotential. The theory is then deformed so that the two or three flat directions at the Mkw point become massive, while the vacuum energy is lifted to a positive value. These theories have a natural separation between the scale of SUSY-breaking (as given by the gravitino mass and the value of the superpotential), and the possibly small lifting to the positive cosmological constant of the dS. 

A possible disadvantage of this scenario is that the same separation of scales also occurs between the gravitino mass and the much smaller uplifted masses of the no-scale directions. A way to remedy this is to turn to supersymmetric Mkw. There, the gravitino mass is small from the start, and the challenge is to break SUSY in such a way that all scalar masses become larger than the gravitino mass, while the cosmological constant remains small.

The generic supersymmetric Mkw is fully stable and lacks flat directions. Unfortunately, as proven in \cite{Kallosh:2014oja}, any small deformation of the superpotential (or the K{\"a}hler potental) just yields another supersymmetric vacuum -- Mkw or AdS. Hence, one can never get a dS in this way. The proposed solution of \cite{Kallosh:2014oja} is to add a new field, referred to as a Polonyi field, to generate an additional, flat direction. The Mkw minimum can then be uplifted to a stable dS. Through appropriate choices of parameters, in \cite{Kallosh:2014oja} it is argued that all scalar masses can be made larger than the gravitino mass, and that the cosmological constant can be set at a still much smaller scale.

A drawback of the non-perturbative models is that the exact form of the corrections is not known in general, and it may become difficult to blindly trust the results. One is therefore motivated to try to do without such non-perturbative effects and generate dS solutions perturbatively. This has turned out to be remarkably complicated in simple string motivated examples, and various no-go results have been obtained in the context of geometric compactifications of type II theories \cite{Maldacena:2000mw,Hertzberg:2007wc,Caviezel:2008tf}. 

The natural models to study are the so called STU-models obtained through compactifying type IIB, type IIA (or M-theory) on twisted tori with orientifolds \cite{Derendinger:2004jn,Dall'Agata:2005fm,Danielsson:2009ff,Dall'Agata:2009gv,Dibitetto:2011gm,Danielsson:2011au,Danielsson:2012et}. These models have an isotropic sector with three complex moduli ($S$, $T$ and $U$), while the full non-isotropic theory has $7$ complex moduli. Different theories are specified through various kinds of fluxes due to the metric and gauge fields living on the manifold. In order to find stable dS solution it has turned out to be crucial to turn on non-geometric fluxes, and explicit, successful examples can be found in refs~\cite{deCarlos:2009fq,Danielsson:2012by,Blaback:2013ht}.

The stable dS in \cite{deCarlos:2009fq,Danielsson:2012by,Blaback:2013ht} were all found to be close to Mkw vacua with no special properties. The aim of our work is to investigate the existence of examples close to either no-scale or fully supersymmetric Mkw. In this paper we will see that this is indeed the case.
In fact, we will be able to show that all vacua generated through the addition of non-perturbative terms can be captured through the presence of non-geometric fluxes. The key to this is the fact that all the relevant properties of the vacua, including the properties of the mass matrix only depend on up to third derivatives of the superpotential. As we will see, the number of real degrees of freedom in the complex superpotential up to third order derivatives is $32$, which coincides with the number of real fluxes in the most general duality invariant superpotential. From this point of view the non-perturbative terms do not introduce anything new beyond the polynomial form.

In case of no-scale, we will give examples of non-perturbative potentials admitting stable dS solutions approaching Mkw points with two as well as three massless directions. In both cases we provide the equivalent polynomial superpotential that gives the same values for the cosmological constant and the masses. We also provide the first example where one of the squared masses go through zero and changes sign at the no-scale point. The existence of such a possibility was argued for in \cite{Marsh:2014nla} but no explicit example was provided. It is crucial that our example is of a more general type than those accessible through the exponential.
 
In case of supersymmetric Mkw, our analysis more or less automatically selects the restricted class of points that have flat directions. Hence, we can find deformations that take us to dS vacua, and with some care these can be made stable. Our construction does not need the introduction of ad hoc Polonyi fields.

The paper is organised as follows. In section 2 we review STU-models arising from type IIB compactifications, and the duality-covariant form of their superpotential induced by generalised fluxes. In section 3 we outline a systematic way of solving the equations of motion, while keeping track of the stability of the theory close to Mkw or no-scale points. In section 4 we have collected all our explicit examples. Section 5 contains our conclusions and some outlook. We also provide two appendices that classify all supersymmetric Mkw points, as well as no-scale points, for the full polynomial superpotential.

\section{Type IIB compactifications with O$3$/O$7$-planes}
\label{sec:Z2xZ2}

Type IIB compactifications on $T^{6}/\left(\mathbb{Z}_{2}\,\times\,\mathbb{Z}_{2}\right)$ with O$3$/O$7$-planes and D$3$/D$7$-branes as well as all dual backgrounds, can be incorporated within a class of $\mathcal{N}=1$ 
supergravity theories \emph{a.k.a.} STU-models. Such 4D effective descriptions enjoy, in their isotropic incarnation, an $\textrm{SL}(2)^{3}$ global symmetry which can be interpreted as string duality 
relating dual ten-dimensional backgrounds to each other.

Isotropic STU-models of this type arise from the coupling between the gravity multiplet and three chiral multiplets. These theories are then free of vectors and possess a scalar sector containing three
complex scalars denoted by $\Psi^{\a}\,\equiv\,\left(S,\,T,\,U\right)$ spanning the $\left(\frac{\textrm{SL}(2)}{\textrm{SO}(2)}\right)^{3}$ coset. Adopting the type IIB language, the $S$ modulus is the one
that contains the ten-dimensional dilaton, whereas the $T$ and $U$ moduli are interpreted as K\"ahler and complex structure moduli respectively. 

The kinetic Lagrangian 
\be
\mathcal{L}_{\textrm{kin}} = \frac{\partial S\partial \overline{S}}{\left(-i(S-\overline{S})\right)^2} \,+ \, 3\,\frac{\partial T\partial \overline{T}̣}{\left(-i(T-\overline{T})\right)^2}\, + \,
 3\,\frac{\partial U\partial \overline{U}}{\left(-i(U-\overline{U})\right)^2} \ ,
\ee
can be derived from the following K\"ahler potential
\be
\label{Kaehler_STU}
K\,=\,-\log\left(-i\,(S-\overline{S})\right)\,-\,3\,\log\left(-i\,(T-\overline{T})\right)\,-\,3\,\log\left(-i\,(U-\overline{U})\right)\ .
\ee
Note that we are adopting a set of conventions where the imaginary parts of the complex fields are those ones carrying geometric meaning and hence they appear in the above K\"ahler potential and
need to be strictly positive. The real parts, on the contrary, just represent axionic scalars and have no sign restriction. As a consequence, our choice for the \emph{origin} of moduli space is given by
\be
\label{origin}
S_{0} \ = \ T_{0} \ = \ U_{0} \ = \ i \ .
\ee

In this paper we will analyse different mechanisms giving rise to scalar potentials for the $\left(S,\,T,\,U\right)$ moduli. Since there are no vector fields available, a potential cannot be induced by
means of a gauging procedure. However, some other massive deformations turn out to be  consistent with minimal supersymmetry. 
Such deformations are controlled by an arbitrary holomorphic function $W(S,T,U)$ called \emph{superpotential}, which induces a scalar potential through
\be
\label{V_N=1}
V\,=\,e^{K}\left(-3\,|W|^{2}\,+\,K^{\a\bar{\b}}\,D_{\a}W\,D_{\bar{\b}}\overline{W}\right)\ ,
\ee
where $K^{\a\bar{\b}}$ is the inverse K\"ahler metric and $D$ denotes the K\"ahler-covariant derivative.
The value of $|W|$ at a given point in moduli space sets the scale of the gravitino mass, whereas the size of the F-terms $|DW|$, which are responsible for supersymmetry breaking sets the mass scale of the
spin-$\tfrac{1}{2}$ fermions. 

There are mainly two distinct ways of inducing a holomorphic superpotential in the context of type II string compactifications: a \emph{perturbative} one and a \emph{non-perturbative} one.
The former includes the possibility of adding \emph{e.g.} gauge fluxes and metric flux. Within the latter class instead, there appear phenomena such as gaugino condensation in the open-string sector or
D-brane instanton effects.

In the context of isotropic STU-models, sets of stable dS critical points have been found both with pertubatively 
\cite{deCarlos:2009fq,Danielsson:2012by,Blaback:2013ht} and non-pertubatively \cite{Danielsson:2013rza,Blaback:2013qza} induced superpotentials. These critical points always turn out to be organised into thin regions in parameter space attached to a line of Mkw solutions. 
Following the strategy of searching for stable dS solutions around Mkw points, it becomes natural to study the possibility of 
deforming special Mkw solutions, \emph{e.g.} of SUSY or no-scale type, into a stable dS vacuum.

This work is inspired by the line of investigation opened up in refs~\cite{Kallosh:2014oja,Marsh:2014nla}. There, the importance of approximate no-scale dS vacua in particular is discussed and a technical machinery is developed for constructing simple analytical families of such solutions. In the construction, the key ingredient seems to be the presence of a non-zero \emph{third} derivative of the superpotential w.r.t. the K\"ahler moduli. However, all explicit realisations shown there, seem to rely on the presence of
non-perturbative effects, whereas the possibility of finding examples within perturbatively-induced superpotentials of polynomial type
still remains to be verified.

The main goal of this paper will be that of testing STU-models with polynomial superpotentials induced by generalised fluxes when it comes to searching for dS solutions allowing for full analytical treatment. We will follow the approach of \cite{Kallosh:2014oja} in order to construct simple examples of stable dS vacua both close to SUSY Mkw and close to no-scale points. All of this will be achieved
 within a string-inspired STU-model with a duality-covariant superpotential.

\subsection*{Perturbatively-induced superpotentials}

In the type IIB duality frame of our interest, gauge fluxes of $F_{(3)}$ \& $H_{(3)}$ type turn out to be allowed by a combination of the orbifold and orientifold involution, other RR gauge fluxes and metric flux in 
completely projected out. This is what identifies the complete set of type IIB geometric fluxes, \emph{i.e.} deformations admitting a 10D understanding.

In the following subsections we will see that superpotentials which are purely induced by gauge fluxes are unable to generate a dependence on the K\"ahler moduli.
%In order to obtain such a dependence in this context, one needs to add generalised fluxes obtained by acting on the geometric ones with a duality chain.
% "one needs" sounds like "it is necessary", which it is not, NP also works.
In order to obtain such a dependence in this context, one can add generalised fluxes obtained by acting on the geometric ones with a duality chain.

\subsubsection*{Models with only gauge fluxes}

Type IIB compactifications with only $F_{(3)}$ \& $H_{(3)}$ fluxes were originally studied in ref.~\cite{Giddings:2001yu}. These backgrounds are supported by the presence of O$3$-planes and D$3$-branes in
 flat geometry. The corresponding flux-induced superpotential reads \cite{Gukov:1999ya}
\be
\label{W_GKP}
W_{\textrm{GKP}}\,=\underbrace{\,a_{0}\,-\,3a_{1}\,U\,+\,3a_{2}U^{2}\,-\,a_{3}\,U^{3}\,}_{F_{(3)} \textrm{ flux}}\,-\,\underbrace{\,S\,\left(b_{0}\,-\,3b_{1}\,U\,+\,3b_{2}U^{2}\,-\,b_{3}\,U^{3}\right)\,}_{H_{(3)} \textrm{ flux}}\ ,
\ee
where all superpotential couplings are consistently chosen to be real thanks to our conventions \eqref{origin} for the origin of moduli space. 

The $\mathcal{N}=1$ supergravity defined by the above superpotential has a so-called \emph{no-scale} symmetry due to the absence of the K\"ahler modulus $T$. This implies that its real and imaginary parts respectively 
appear as a completely flat and a \emph{run-away} direction in the scalar potential. The latter in turn implies that the only maximally symmetric solutions that these models can have must be Minkowski.

\subsubsection*{Models with generalised fluxes}

Starting from a geometric STU-model, one can start acting with $\textrm{SL}(2)^{3}$ transformations to obtain dual models. In this way, it becomes natural to conjecture the existence of a completely 
duality-covariant superpotential \cite{Shelton:2005cf} containing all possible STU-terms up to linear in $S$ and up to cubic in $T$ \& $U$.
We will now summarise here the correspondence between generalised isotropic fluxes and superpotential couplings appearing in the $\mathcal{N}=1$ effective 4D description. 
The complete generalised flux-induced superpotential can be written as 
\be
\label{W_all_fluxes}
W_{\textrm{pert.}} = (P_{F} - P_{H} \, S ) + 3 \, T \, (P_{Q} - P_{P} \, S ) + 3 \, T^2 \, (P_{Q'} - P_{P'} \, S ) + T^3 \, (P_{F'} - P_{H'} \, S ) \ ,
\ee
where\footnote{\label{c1_tilde}Please note that, in principle, the truncation to the isotropic sector gives rise to $32+8=40$ fluxes, where all the fluxes transforming in the mixed symmetry representations of $\textrm{GL}(6)$ (\emph{i.e.} $Q$, $P$ and their primed counterparts) have in fact two fluxes $(c_{1},\tilde{c}_{1})$ etc. giving rise to one single coupling $(2c_{1}-\tilde{c}_{1})$ etc., so without loss of generality, we set throughout the text $\tilde{c}_{1}=c_{1}$ etc..}
the couplings in
\be
\begin{array}{lcll}
\label{Poly_unprim}
P_{F} = a_0 - 3 \, a_1 \, U + 3 \, a_2 \, U^2 - a_3 \, U^3 & \hspace{5mm},\hspace{5mm} & P_{H} = b_0 - 3 \, b_1 \, U + 3 \, b_2 \, U^2 - b_3 \, U^3 & ,  \\[2mm]
P_{Q} = c_0 + c_{1} \, U - c_{2} \, U^2 - c_3 \, U^3 & \hspace{5mm},\hspace{5mm} & P_{P} = d_0 + d_{1} \, U - d_{2} \, U^2 - d_3 \, U^3 & ,
\end{array}
\ee
are introduced and explained in table~\ref{table:unprimed_fluxes}, whereas the details of the couplings in
\be
\begin{array}{lcll}
\label{Poly_prim}
P_{F'} = a_3' + 3 \, a_2' \, U + 3 \, a_1' \, U^2 + a_0' \, U^3 & \hspace{3mm},\hspace{3mm} & P_{H'} = b_3' + 3 \, b_2' \, U + 3 \, b_1' \, U^2 + b_0' \, U^3 & ,  \\[2mm]
P_{Q'} = -c_3' +  c'_{2} \, U + c'_{1} \, U^2 - c_0' \, U^3 & \hspace{3mm},\hspace{3mm} & P_{P'} = -d_3' + d'_{2} \, U + d'_{1} \, U^2 - d_0' \, U^3 & ,
\end{array}
\ee
are given in table~\ref{table:primed_fluxes}.
The first half of the terms (see table~\ref{table:unprimed_fluxes}) are characterised by lower powers in $T$, \emph{i.e.} up to linear, and represent fluxes which admit a locally geometric interpretation in type IIB (unprimed fluxes). 
The remaining ones (primed fluxes) instead, appear with quadratic and cubic behaviour in $T$ and represent additional generalised fluxes which do not even admit a locally geometric description. These were first formally introduced in ref.~\cite{Aldazabal:2006up} as dual counterparts of the unprimed fluxes. 

\begin{table}[h!]
\renewcommand{\arraystretch}{1.25}
\begin{center}
\scalebox{0.92}[0.92]{
\begin{tabular}{ | c || c | c |}
\hline
couplings & Type IIB &  fluxes \\
\hline
\hline
$1 $&  $ {F}_{ ijk} $&  $  a_0 $ \\
\hline
$U $&  ${F}_{ ij c} $&  $   a_1 $ \\
\hline
$U^2 $& ${F}_{i b c} $&  $  a_2 $\\
\hline
$U^3 $& ${F}_{a b c} $& $  a_3 $ \\
\hline
\hline
$S $& $ {H}_{ijk} $& $  -b_0$ \\
\hline
$S \, U $& ${H}_{ij c} $&  $  -b_1 $ \\
\hline
$S \, U^2 $&  ${H}_{ i b c}$   & $  -b_2 $ \\
\hline
$S \, U^3 $& $ {H}_{a b c} $&  $  -b_3 $ \\
\hline
\hline
$T $& $  {Q_k}^{a b} $& $  c_0 $ \\
\hline
$T \, U $& $ {Q_k} ^{a j} = {Q_k}^{i b} \,\,\,,\,\,\, {Q_a}^{b c} $&  $c_1 \,\,\,,\,\,\, \tilde {c}_1 $ \\
\hline
$T \, U^2 $& $ {Q_c}^{ib} = {Q_c}^{a j} \,\,\,,\,\,\, {Q_k}^{ij} $ & $c_2 \,\,\,,\,\,\,\tilde{c}_2 $ \\
\hline
$T \, U^3 $& $  {Q_{c}}^{ij} $&  $c_3 $ \\
\hline
\hline
$S \, T $& $ {P_k}^{a b}$ &  $  -d_0 $ \\
\hline
$S \, T \, U $& $ {P_k}^{a j} = {P_k}^{i b} \,\,\,,\,\,\, {P_a}^{b c} $&   $-d_1 \,\,\,,\,\,\, -\tilde{d}_1 $ \\
\hline
$S \, T \, U^2 $& $ {P_c}^{ib}= {P_c}^{a j} \,\,\,,\,\,\, {P_k}^{ij} $&  $-d_2 \,\,\,,\,\,\,-\tilde{d}_2 $\\
\hline
$S \, T \, U^3 $& $  {P_{c}}^{ij} $&   $-d_3 $ \\
\hline
\end{tabular}
}
\end{center}
\caption{{\it Mapping between unprimed fluxes and couplings in the superpotential in type IIB with O3 and O7. The six internal directions are split into $\,``-"$ labelled by $i=1,3,5$ and $\,``\,|\,"$ labelled by $a=2,4,6$.}}
\label{table:unprimed_fluxes}
\end{table}

\begin{table}[h!]
\renewcommand{\arraystretch}{1.25}
\begin{center}
\scalebox{0.92}[0.92]{
\begin{tabular}{ | c || c | c |}
\hline
couplings &  Type IIB &  fluxes \\
\hline
\hline
$T^3 \, U^3 $& $ {F'}^{ijk} $&  $  a_0' $ \\
\hline
$T^3 \, U^2 $& ${F'}^{ ij c} $& $   a_1' $ \\
\hline
$T^3 \, U $& ${F'}^{i b c} $& $  a_2' $ \\
\hline
$ T^3 $& ${F'}^{a b c} $& $  a_3' $ \\
\hline
\hline
$S \, T^3 \, U^3 $& $ {H'}^{ ijk} $& $  -b_0'$ \\
\hline
$S \, T^3 \, U^2 $& $ {H'}^{i jc} $& $ - b_1' $ \\
\hline
$S \, T^3 \, U $& $ {H'}^{ i b c} $& $  -b_2' $ \\
\hline
$S  \, T^3 $& $ {H'}^{a b c} $& $  -b_3' $ \\
\hline
\hline
$T^2 \, U^3 $& $  {{Q'}_{a b}}^k $& $  c_0' $\\
\hline
$T^2 \, U^2 $& $ {{Q'}_{a j}}^k = {{Q'}_{i b}}^k \,\,\,,\,\,\, {{Q'}_{b c}}^a $& $c_1' \,\,\,,\,\,\, \tilde{c}_1' $\\
\hline
$T^2 \, U $& $ {{Q'}_{ib}}^c = {{Q'}_{a j}}^c \,\,\,,\,\,\, {{Q'}_{ij}}^k $& $c_2' \,\,\,,\,\,\,\tilde{c}_2' $ \\
\hline
$T^2 $& $ {{Q'}_{ij}}^{c} $ &$c_3' $ \\
\hline
\hline
$S \, T^2 \, U^3$& $  {{P'}_{a b}}^k $ &$  -d_0' $ \\
\hline
$S \, T^2 \, U^2 $& $ {{P'}_{a j}}^k = {{P'}_{i b}}^k \,\,\,,\,\,\, {{P'}_{b c}}^a $ & $-d_1' \,\,\,,\,\,\, -\tilde{d}_1' $ \\
\hline
$S \, T^2 \, U $& $ {{P'}_{ib}}^c = {{P'}_{a j}}^c \,\,\,,\,\,\, {{P'}_{ij}}^k $& $-d_2' \,\,\,,\,\,\,-\tilde{d}_2' $ \\
\hline
$S \, T^2  $& $  {{P'}_{ij}}^{c} $&  $-d_3' $\\
\hline
\end{tabular}
}
\end{center}
\caption{{\it Mapping between primed fluxes and couplings in the superpotential. The conventions are the same as in table~\protect\ref{table:unprimed_fluxes}.}}
\label{table:primed_fluxes}
\end{table}

\subsection*{Non-perturbatively-induced superpotentials}

Starting again from the geometric superpotential \eqref{W_GKP} induced by gauge fluxes, one could instead introduce non-perturbative effects in order to generate a $T$-dependence in the superpotential, 
which is typically exponential. Following the idea of \cite{Saltman:2004sn}, these could be used in order to further fix $T$ in perturbatively-constructed dS solutions where it appears as a \emph{run-away} direction.
Such non-perturbative effects can thus be seen as an alternative way of breaking the no-scale symmetry of \eqref{W_GKP} w.r.t. generalised fluxes. 

As mentioned earlier, amongst possible mechanisms to generate this type of superpotentials, we find the phenomenon of gaugino condensation within the gauge theory sector living on D$7$-branes or D-brane 
instanton effects.
In particular, the line of including gaugino condensation \cite{Font:1990nt} has been considered in the literature as a possible mechanism to further stabilise the K\"ahler moduli (see \emph{e.g.} 
refs~\cite{Witten:1996bn, Achucarro:2006zf}).

A fairly generic prototype of non-perturbatively-induced superpotential can be written as
\be
\label{W_nonpert}
W_{\textrm{non-pert.}} = \underbrace{\,(P_{F} - P_{H} \, S )\,}_{W_{\textrm{GKP}}} \, + \, P_{Z} \, e^{i\,\a T} \ ,
\ee 
where $P_{Z}$ is a priori an arbitrary holomorphic function of $S$ \& $U$ and $\a>0$ is some characteristic constant that depends on the physics of the explicit non-perturbative phenomenon in question
\footnote{In the case of gaugino condensation, a rough estimation for $\a$ is $\frac{2\pi}{N}$, $N$ being the rank of the corresponding $\textrm{SU}(N)$ gauge group.}. 
Unfortunately, not much is known about the explicit form of the function $P_{Z}(S,U)$ since it would be extremely difficult to compute from stringy principles. In the construction of 
ref.~\cite{Kachru:2003aw}, an argument was presented that would allow one to ignore such an $(S,U)$-dependence in $P_{Z}$, thus reducing the prefactor in front of the exponential to a constant, at least
in the large volume regime.

\section{Solving the field equations systematically}
\label{sec:method}

In order to solve the field equations for the six real scalars in the STU-model of our interest and find maximally symmetric solutions, we combine two crucial observations that transform the extremality 
conditions for the scalar potential \eqref{V_N=1} into an algebraic system of linear equations in the superpotential couplings. 

The first fact that we use is that a general non-compact $\textrm{SL}(2)^{3}$ duality transformation takes any point in moduli space to the origin \eqref{origin}. This statement holds for any supergravity
model in which the scalars span a homogenous space, just as \emph{e.g.}, $\left(\frac{\textrm{SL}(2)}{\textrm{SO}(2)}\right)^{3}$. Due to the general form of the scalar potential, the formulation of its 
extremality conditions in the origin is just given by a set of six algebraic quadratic equations in the superpotential couplings. 

Not only does this simplify the problem significantly, but such a restricted search for critical points can even turn out to be exhaustive if one moreover includes a set of superpotential terms which 
happens to be closed under non-compact duality transformations \cite{Dibitetto:2011gm}. In our case, for a pertubatively-induced superpotential, this is in particular true if one keeps the complete
duality-invariant superpotential given in \eqref{W_all_fluxes}.

It may be worth mentioning that, for the case of non-pertubatively induced superpotentials of the type in \eqref{W_nonpert}, such a search for solutions will generically imply a loss of generality.
This is specifically related to the fact that the $\textrm{SL}(2)_{T}$ symmetry appears to be broken by the presence of the exponential term \eqref{W_nonpert}, unless a non-trivial compensating transformation
under $\textrm{SL}(2)_{T}$ is allowed for the constant $\a$ appearing there.

The second prescription that we make use of, is that of treating the derivatives of the superpotential w.r.t. the three complex fields
as independent quantities. This line of investigation was first pursued in ref.~\cite{Danielsson:2012by} where the so-called \emph{SUSY-breaking parameters} were introduced. By looking at the form of the F-terms in the origin of moduli space,
\be
F_{\a} \ \equiv \ \left. D_{\a}W \right|_{\textrm{origin}} \ = \ \textrm{constants} \ ,
\ee
one can reexpress six of the fluxes as functions of the above constants. Upon doing so, one can use the remaining fluxes to easily solve the field equations, since they can appear there at most linearly.

However, it was later noted in ref.~\cite{Kallosh:2014oja} that such a method is nothing but an overconstrained case of the more general situation where all derivatives of the superpotential evaluated in the origin up to third order take part in a various-step-procedure. Specifically, if one starts out with the following arbitrary cubic superpotential\footnote{Such a form of the superpotential might even be supported by geometric arguments, like \emph{e.g.} those in ref.~\cite{Catino:2013ppa}, where a cubic term of this form was proposed in order to reproduce truncations of some particular gaugings in maximal supergravity describing relevant M-theory compactifications including the $7$-sphere.} 
\be
\label{W_Phi}
W\left(\Phi^{\a}\right) \ = \ W_{0} \ + \ W_{\a} \, \Phi^{\a} \ + \ \frac{1}{2!} \, W_{\a\b} \, \Phi^{\a}\Phi^{\b} \ + \ \frac{1}{3!} \, W_{\a\b\g} \, \Phi^{\a}\Phi^{\b}\Phi^{\g} \ ,
\ee
where $\Phi^{\a} \ \equiv \ \left(S-i,\,T-i,\,U-i\right)$ and all $W$ derivatives appear as arbitrary complex numbers, one can:

\begin{itemize}
\item Choose $W_{0}$ in order to fix the gravitino mass scale,
\item Choose $W_{\a}$ in order to fix the SUSY-breaking scale,
\item Fix part of the $W_{\a\b}$'s in order to solve the field equations (where they only appear \emph{linearly}), while suitably choosing the remaining ones,
\item Fix $W_{\a\b\g}$ such that the mass matrix (where, as well, they only appear \emph{linearly}) be positive definite.
\end{itemize}

The mass matrix for the six real moduli fields at a critical point has the following general form
\be
{\left(m^{2}\right)^{I}}_{J} \ = \ \left(
\begin{array}{cc}
K^{\a\bar{\g}}\,D_{\bar{\g}}D_{\b}V & K^{\a\bar{\g}}\,D_{\bar{\g}}D_{\bar{\b}}V \\
K^{\bar{\a}\g}\,D_{\g}D_{\b}V & K^{\bar{\a}\g}\,D_{\g}D_{\bar{\b}}V
\end{array}\right) \ = \ \left(
\begin{array}{cc}
K^{\a\bar{\g}}\,V_{\bar{\g}\b} & K^{\a\bar{\g}}\,V_{\bar{\g}\bar{\b}} \\
K^{\bar{\a}\g}\,V_{\g\b} & K^{\bar{\a}\g}\,V_{\g\bar{\b}}
\end{array}\right) \ ,
\ee
where $V_{\a\b}$ etc. just denote ordinary derivatives of $V$ w.r.t. the complex fields $\Psi^{\a}$. In order to significantly simplify the problem of getting all positive eigenvalues and hence construct
proper minima of the potential, it might be very useful in some cases to observe \cite{Kallosh:2014oja} that the mass matrix becomes block-diagonal upon imposing $V_{\a\b}\,=\,0$. Such a condition, which
can be easily solved in terms of the $W_{\a\b\g}$, also turns out to enforce a pairwise organisation of the mass spectrum, which will in such cases only possess three distinct eigenvalues.

The positivity of the cosmological constant, instead, is here controlled by the sign of the following quadratic combination
\be
V_{0} \ \propto \ -3\,\left|W_{0}\right|^2 \ + \ \left|W_{\a} \, + \, K_{\a}W_{0}\right|^2 \ ,
\ee
which means that such a procedure can systematically produce analytical stable dS solutions provided that all the different complex coefficients in the \eqref{W_Phi} can be fixed independently.

Due to the form of the K\"ahler potential \eqref{Kaehler_STU}, we should restrict to those complex coefficients that do not generate
superpotential terms with degree higher than one w.r.t. $S$, \emph{i.e.} $W_{SS}\,=\,W_{SSS}\,=\,W_{SST}\,=\,W_{SSU}\overset{!}{=}\,0$.
This implies that the most general parametrisation of the form \eqref{W_Phi} compatible with the class of STU-models presented in section~\ref{sec:Z2xZ2} counts $16$ complex parameters, of which $1$ is given by $W_{0}$, $3$ by $W_{\a}$, $5$ by $W_{\a\b}$ and the remaining $7$ by $W_{\a\b\g}$.

The final crucial observation is now that such a parameter space is in \emph{one-to-one} correspondence with the complete set of $32$ real superpotential couplings collected in tables~\ref{table:unprimed_fluxes} and \ref{table:primed_fluxes} describing the most general duality-invariant superpotential for our STU-model. The mapping relating generalised fluxes to complex superpotential derivatives is, in fact, linear and invertible. This in particular implies that, whenever a stable dS solution is found for a certain superpotential derivative configuration, this will always admit an STU-realisation in terms of $32$ generalised perturbative fluxes.

\subsection*{Stable dS vacua close to SUSY Mkw points}

Let us now apply the above prescription in order to find interesting dS vacua obtained by starting from a SUSY Mkw solution and subsequently breaking SUSY by small amounts. The details concerning the most general supersymmetric Mkw solution within our STU-model can be found in appendix~\ref{app:SUSY_Mkw}. 

In order to apply the above formalism, we first need to understand our Mkw points in the language of $W$ derivatives. These are identified by the following choice:
\be \label{NoScalePointConditions}
\begin{array}{lcclclclcc}
W_{0} \, = \, 0 & , & & \textrm{and} & & W_{\a} \, = \, 0 & , & \a\,=\,S,\,T,\,U & & ,
\end{array}
\ee
whereas all the other $12$ complex higher derivatives stay completely arbitrary. One can check that the above choice already automatically implies the field equations and guarantees the 
(semi-)positiveness of the mass matrix.

Now we need to construct a consistent \emph{Ansatz} to break SUSY by small amounts and hence move away from the SUSY Mkw critical point. To this end, we introduce a small parameter $\epsilon$, \emph{i.e.}
such that $|\epsilon|\,\ll\,1$, that can be taken to zero whenever one wants to go back to the supersymmetric situation. Here we can make use of the no-go theorem in ref.~\cite{Kallosh:2014oja}, which guarantees that our construction automatically selects points close to a SUSY Mkw with two real massless directions. Without flat directions the theorem only allows for deformations into other SUSY points, and any dS critical point is excluded. 

Such a deformation \emph{Ansatz} reads
\be
\label{Close2SUSY}
\begin{array}{lcclclcl}
W_{0} \, = \, \kappa_{0} \, \epsilon   & , & & \textrm{and} & & W_{\a} \, = \, \kappa_{\a} \, \epsilon & ,
\end{array}
\ee
where $\left(\kappa_{0},\,\kappa_{S},\,\kappa_{T},\,\kappa_{U}\right)$ are arbitrary complex numbers of $\mathcal{O}(1)$. By plugging the \eqref{Close2SUSY} into the field equations, the second 
derivatives of $W$ can only appear there linearly. Hence, one can use six out of the ten real independent parameters in $W_{\a\b}$ in order to find extrema of the potential. The other four of them can
be arbitrarily chosen.

As far as the mass matrix is concerned, the remarkable feature of these dS points close to a SUSY Mkw with two flat directions is that the light directions always coincide with the two real sGoldstini
\be
g_{\a} \ \equiv \ \frac{F_{\a}}{\sqrt{F_{\b}\,\overline{F}^{\b}}} \ ,
\ee
\emph{i.e.} the superpartners of the Goldstone fermions responsible for the SUSY-breaking mechanism. This implies that these are the only directions that can turn into tachyons in the small $\epsilon$ 
limit.

At this point it is useful to impose the pairwise degeneracy condition for the mass spectrum, \emph{i.e.} $V_{\a\b}\,=\,0$, in order for the two sGoldstini directions to become degenerate. Such a situation turns out to be very
special since it produces two sGoldstini directions with the same mass given by their \emph{average}, which happens to be subject to a universal geometric stability bound \cite{Covi:2008cn,Borghese:2012yu}
\be
\label{sG_bound}
\eta_{sG} \ \equiv \ \frac{1}{V} \, g^{\bar{\a}}\bar{g}^{\b} \, D_{\bar{\a}}D_{\b}V \ = \ \frac{2}{3\g} \ - \ \frac{1+\g}{\g} \, \tilde{\mathcal{R}} \  \overset{!}{\geq} \ 0 \ ,
\ee
where $\g \, \equiv \, \frac{V}{3\,e^{K}\,|W|^{2}}$ and $\tilde{\mathcal{R}} \, \equiv \, \mathcal{R}_{\bar{\a}\b\bar{\g}\d} \, g^{\bar{\a}}\bar{g}^{\b}g^{\bar{\g}}\bar{g}^{\d}$ 
denotes the sectional curvature of the K\"ahler manifold along the $g_{\a}$ plane.
As already noted in ref.~\cite{Danielsson:2012by}, the sectional curvature can be rewritten as $\tilde{\mathcal{R}}\,=\,\frac{2}{n_{\textrm{eff}}}$, where
\be
n_{\textrm{eff}} \ = \ \frac{\left(F_{\a}\overline{F}^{\a}\right)^{2}}{\left(F_{S}\overline{F}^{S}\right)^{2}\,+\,\frac{1}{3} \, \left(F_{T}\overline{F}^{T}\right)^{2} 
\,+\,\frac{1}{3} \, \left(F_{U}\overline{F}^{U}\right)^{2}} \ ,
\ee
effectively represents the number of complex fields\footnote{The quantity $n_{\textrm{eff}}$ is only defined when SUSY is broken and it ranges from $1$ to $7$. 
The field $S$ has weight one, whereas $T$ \& $U$ have weight three due to the origin of this STU-model as the isotropic limit of the $\textrm{SL}(2)^{7}$ model.} taking part in the SUSY-breaking 
mechanism. As an interesting consequence, the geometric bound in \eqref{sG_bound} only depends on information encoded in the zero-th \& first derivatives of the superpotential, \emph{i.e.} $W_{0}$ \& $W_{\a}$,
 which are parametrised by $\left(\kappa_{0},\,\kappa_{S},\,\kappa_{T},\,\kappa_{U}\right)$ as in \eqref{Close2SUSY} close to a SUSY Mkw point.

Summarising, one can adopt the following combination of a system of equalities and a system of inequalities
\be \label{SUSYIneqs}
\begin{array}{ccccc}
\left\{\begin{array}{lc}
V_{\a} \ \overset{!}{=} \ 0 & , \\
V_{\a\b} \ \overset{!}{=} \ 0 & ,
\end{array}\right.
& & \textrm{and} & & \left\{\begin{array}{lc}
\g \ \overset{!}{>} \ 0 & , \\
n_{\textrm{eff}} \ \overset{!}{>} \ 3\,(1\,+\,\g) & ,
\end{array}\right.
\end{array}
\ee
as a constructive procedure in order to analytically produce simple examples of stable dS vacua close to a SUSY Mkw critical point. We remind the reader that the two sets of equations on the left can be
both linearly solved by fixing six real parameters within $W_{\a\b}$ and ten real parameters in $W_{\a\b\g}$, respectively. The inequalities on the right, can instead be satisfied by suitably choosing
$\kappa_{0}$ \& $\kappa_{\a}$'s. It is worth mentioning that the whole construction still leaves $4+4=8$ real free parameters within the second and third $W$ derivatives.

In section~\ref{sec:examples}, we will present an explicit example of dS vacuum close to a SUSY Mkw point within our class of STU-models inspired by type IIB compactifications with generalised fluxes, which was obtained by following the above procedure.

\subsection*{Stable dS vacua close to no-scale Mkw points}

Another simple way of constructing Mkw points stable up to flat directions, is to consider the so-called 
no-scale models, where SUSY is broken by large amounts but only in a single direction (usually $T$ in our notation) in such a way that
the corresponding F-term contribution to the potential exactly cancels the negative contribution from $-3|W|^2$.
 
A realisation of a no-scale supergravity model is given by the GKP superpotential \eqref{W_GKP} arising from type IIB compactifications
with NS-NS and R-R 3-form gauge fluxes. The possibility of obtaining stable dS solutions in this context was first considered in
 ref.~\cite{Covi:2008ea} and later extensively studied in refs~\cite{Kallosh:2014oja,Marsh:2014nla}, where explicit examples of analytical stable dS solutions close to no-scale Mkw points are constructed by means of non-perturbative superpotentials of the type
\eqref{W_nonpert}.

These examples making use of non-perturbative effects are special realisations of superpotentials within the class given in 
\eqref{W_Phi} with non-trivial zero-th, first, second and third derivatives. Due to the correspondence shown in the beginning of this
 section, they all have a realisation within perturbative (polynomial) superpotentials with generalised fluxes.
In the next section, by making use of this correspondence, we will be able to present stable dS examples close to no-scale points with both two and three massless directions, in the context of our STU-model with $32$ generalised fluxes. 

In order to go beyond the pre-existing examples borrowed from non-perturbative models and construct new solutions close to no-scale Mkw
points from scratch, in analogy with the SUSY case, we need to understand no-scale points in the language of $W$ derivatives.
No-scale Mkw points are identified by
\be
\label{Close2NoScale}
\begin{array}{lc}
\begin{array}{lcccl}
 W_{0} \, = \, \textrm{arbitrary}  &  & , & &  
\left\{\begin{array}{l}
W_{S} \, = \, - K_{S} \, W_{0} \\
W_{U} \, = \, - K_{U} \, W_{0}
\end{array}\right. \end{array} & , \\[6mm] 
\left\{\begin{array}{l}
W_{T} \, = \, 0   \\
W_{ST} \, = \, W_{TU} \, = \, W_{TT} \, = \, 0 \\
W_{STU} \, = \,W_{TUU} \, = \, W_{STT} \, = \, W_{TTU} \, = \, W_{TTT} \, = \, 0
\end{array}\right.
 & ,  \\[3mm]
\end{array}
\ee
whereas all the other complex derivatives stay completely aribitrary. One can check that the above choice already automatically implies the field equations and guarantees the semi-positiveness of the mass matrix. Generically these Mkw solutions will have two
flat directions, but upon imposing a degeneracy condition on the remaining free parameters, one will hit special no-scale points with an extra massless mode. %JB: Since this extra massless mode is of particular interest, maybe when we mention it we should refer to the section "Stable dS close to no-scale", where an example of this thing happening is considered?

By translating the conditions in \eqref{Close2NoScale} into the language of generalised fluxes, one gets easily convinced of the existence of \emph{generalised no-scale points}, \emph{i.e.} for which the explicit flux-induced superpotential may in general depend
on $T$, but in such a way that all its $T$ derivatives up to third order vanish. An exhaustive classification of such models is presented in appendix~\ref{app:no-scale_Mkw}.  

Now we can move away from the generalised no-scale points defined by the conditions in \eqref{Close2NoScale} by deforming them into
the following \emph{Ansatz} for the first $W$ derivatives
\be \label{Close2NSWs}
\left\{\begin{array}{lc}
W_{S} \, = \, - K_{S} \, W_{0} \ + \ \kappa_{S} \, \epsilon & ,\\
W_{T} \, = \, \kappa_{T} \, \epsilon  & ,\\
W_{U} \, = \, - K_{U} \, W_{0} \ + \ \kappa_{U} \, \epsilon & ,
\end{array}\right.
\ee
where $\epsilon$ is again a small real parameter and $\left(\kappa_{S},\,\kappa_{T},\,\kappa_{U}\right)$ are arbitrary 
$\mathcal{O}(1)$ complex parameters. By fixing some of the $W_{\a\b}$'s in order to solve the field equations lineraly, one correctly
approaches a generalised no-scale point with two light directions that are generically lifted at quadratic order in $\epsilon$, just
like in those solutions which are obtained from adding non-perturbative effects. In this situation, fixing some $W_{\a\b\g}$'s to solve
the condition $V_{\a\b} \ \overset{!}{=} \ 0$ can be very helpful in finding stable dS critical points.
 
On the other hand, in order to approach a generalised no-scale Mkw with three flat directions, it turns out that one needs to violate
such a condition since having an odd number of massless modes is not compatible with the pairwise organisation of the mass spectrum
that it causes. In the next section we will show an example of a dS solution exhibiting three light modes, two of which are lifted
at $\mathcal{O}(\epsilon^2)$ while the extra one is lifted at $\mathcal{O}(\epsilon)$, thus providing the first realisation of the
general situation discussed in ref.~\cite{Marsh:2014nla}.

\section{Relevant Examples}
\label{sec:examples}

In this section we will present explicit analytical examples of (stable) dS solutions obtained by making use of the technical machinery presented in section~\ref{sec:method}.
We will first start by giving a generalised-flux realisation of stable dS close to a SUSY Mkw; in this case we will have two light modes lifted at $\mathcal{O}(\epsilon^{2})$. 
Secondly, we will move to discussing dS solutions close to no-scale Mkw points. There we will present both generalised-flux realisations of solutions borrowed from non-perturbative superpotentials 
and new examples which cannot be interpreted as coming from such exponential superpotentials. Only within these new cases will we be able to find the first realisation of an approximate no-scale
dS solution with three light modes, one of which is lifted at linear level, while the other two get lifted at quadratic level. However, such a solution will turn out to be unstable.

\subsection*{Stable dS close to a SUSY Mkw point}

This is a simple example of the application of the constraints in (\ref{SUSYIneqs}). A particularly simple choice that reduces the number of involved real parameters from 6 to 1 is

\be \label{SUSYExFs}
F_{\a} \, = \, \big( \,  (1+i) \, \epsilon  \, ,  \, (\lambda_T +i)  \, \epsilon  \,  ,  \,  (1+i)  \, \epsilon \,  \big) ,
\ee

where $\lambda_T$ can then be seen as a relative scaling between $\mathrm{Re}[F_{T}]$ and $\mathrm{Re}[F_{S}]$ or  $\mathrm{Re}[F_{U}]$. In the language of the prescription given in (\ref{Close2SUSY}), we may additionally pick $\mathcal{O}(\epsilon)$ terms for $W_0$ and write

\be
W_{0} \, = \, \left(\frac{8}{3}+\frac{2 i}{3}\right) \epsilon \ \ \ , \ \ \  W_{S} \, = \, \left(\frac{4}{3}-\frac{i}{3}\right) \epsilon \ \ \ , \ \ \  W_{T} \, = \, \left( \lambda_T+1-3 i\right) \epsilon \ \ \ , \ \ \  W_{U} \, = \, (2-3 i) \epsilon  \, .
\ee

Hence the cosmological constant takes the form

\be
V \, = \, \frac{1}{96} \left(\lambda_T^2-8\right) \, \epsilon^2 \,.
\ee

This two-parameter set of solutions can then be run through the inequalities in (\ref{SUSYIneqs}) which at $\mathcal{O}(\epsilon^0)$ provide the simple bounds

\be \label{SUSYbounds}
-\sqrt{\frac{1}{2} \left(15+\sqrt{385}\right)}< \lambda_T<-2 \sqrt{2} \,\,\,\,\,\, \mathrm{or}\,\,\,\,\,\, 2 \sqrt{2}< \lambda_T<\sqrt{\frac{1}{2} \left(15+\sqrt{385}\right)}
\ee

\begin{table}
\renewcommand{\arraystretch}{1.25}
\begin{center}
\scalebox{0.92}[0.92]{
\begin{tabular}{ | c | c |}
\hline
flux labels &  flux values\\
\hline
\hline
$ a_0 $ & $ \frac{347}{39304}-\frac{12859}{235824} \epsilon $ \\
\hline
$ a_1 $ & $  -\frac{231}{9826}+\frac{25591}{235824}\epsilon $ \\
\hline
$ a_2 $ & $  \frac{407}{29478}+\frac{33905}{235824} \epsilon $ \\
\hline
$ a_3 $ & $ -\frac{623}{39304}-\frac{18281}{235824}\epsilon $ \\
 \hline
\hline
$ b_0 $ & $ -\frac{2029}{39304}-\frac{19429}{78608}\epsilon  $ \\
\hline
$ b_1 $ & $  \frac{13945}{353736}+\frac{5311}{235824}\epsilon  $ \\
\hline
$ b_2 $ & $  -\frac{2455}{176868}-\frac{18209}{78608}\epsilon  $ \\
\hline
$ b_3 $ & $  \frac{1349}{9826}+\frac{10481}{78608}\epsilon  $ \\
\hline
\hline
$ c_0 $ & $  -\frac{3749}{58956} -\frac{37007}{78608} \epsilon $ \\
\hline
$ c_1 $ & $  -\frac{9661}{117912} -\frac{158563}{235824}\epsilon $ \\
\hline
$ c_2 $ & $  \frac{4241}{353736}-\frac{12115}{78608}\epsilon  $ \\
\hline
$ c_3 $ & $ \frac{2569}{58956}-\frac{143}{235824} \epsilon  $ \\
 \hline
\hline
$ d_0 $ & $ \frac{15107}{176868} +\frac{45635}{235824}\epsilon $ \\
\hline
$ d_1 $ & $  -\frac{1625}{58956} +\frac{3083}{78608}\epsilon $ \\
\hline
$ d_2 $ & $  -\frac{20293}{117912} -\frac{10159}{78608}\epsilon$ \\
\hline
$ d_3 $ & $ \frac{11795}{353736} +\frac{11821}{235824}\epsilon $ \\
\hline
\end{tabular} \quad
\begin{tabular}{ | c | c |}
\hline
flux labels &  flux values\\
\hline
\hline
$ a_0' $ & $ \frac{1333}{39304}+\frac{20613}{78608}\epsilon $ \\
\hline
$ a_1' $ & $ \frac{6943}{176868}+\frac{78155}{235824}\epsilon  $ \\
\hline
$ a_2' $ & $ -\frac{4307}{176868} +\frac{15851}{235824}\epsilon $ \\
\hline
$ a_3' $ & $ \frac{23869}{117912} +\frac{253991}{235824}\epsilon $ \\
\hline
\hline
$ b_0' $ & $ \frac{133}{19652}+\frac{6347}{235824}\epsilon  $ \\
\hline
$ b_1' $ & $ \frac{110}{14739}-\frac{10311}{78608}\epsilon  $ \\
\hline
$ b_2' $ & $ -\frac{8225}{353736} -\frac{32561}{235824}\epsilon $ \\
\hline
$ b_3' $ & $ \frac{3441}{39304}+\frac{119707}{235824}\epsilon  $ \\
\hline
\hline
$ c_0' $ & $ \frac{2681}{88434}+\frac{7843}{235824}\epsilon  $ \\
\hline
$ c_1' $ & $ \frac{43471}{353736}+\frac{95315}{235824}\epsilon  $ \\
\hline
$ c_2' $ & $ -\frac{32513}{353736} -\frac{46999}{78608}\epsilon $ \\
\hline
$ c_3' $ & $ -\frac{1691}{14739} -\frac{27649}{235824}\epsilon $ \\
\hline
\hline
$ d_0' $ & $ -\frac{1075}{117912} -\frac{9989}{235824}\epsilon $ \\
\hline
$ d_1' $ & $ \frac{32869}{353736}-\frac{12817}{235824}\epsilon  $ \\
\hline
$ d_2' $ & $ -\frac{1370}{14739} -\frac{66721}{235824}\epsilon $ \\
\hline
$ d_3' $ & $ -\frac{2578}{44217} -\frac{23815}{78608}\epsilon $ \\
\hline
\end{tabular}
}
\end{center}
\caption{{\it An explicit example of a generalised flux realisation of a stable dS solution which analytically approaches a SUSY Mkw point as $\epsilon\rightarrow 0$. It may be checked explicitly that the 0-th order values of the fluxes define a SUSY Mkw solution with two flat directions.}}
\label{table:SUSYExFluxes}
\end{table}

The equations in (\ref{SUSYIneqs}) can be solved to fix $W_{ST}$, $W_{TT}$ and $W_{TU}$ (with $V_{\alpha} = 0$) as well as $W_{STT}$,  $W_{STU}$, $W_{TTT}$, $W_{TTU}$ and $W_{TUU}$ (with $V_{\alpha \beta} = 0$). Then it only remains to choose values for the 8 remaining free real parameters. Here we will present one particular realization of fluxes obtained with
\be
W_{SU} \, = \, \frac{1}{3}\left(1+i\right) \ \ \ , \ \ \  W_{UU} \, = \, \frac{1}{3}\left(1+i\right) \ \ \ , \ \ \  W_{SUU} \, = \, W_{UUU} = 0  \, .
\ee

\begin{figure}
  \centering
       \includegraphics[bb=0 0 7in 6.32in,keepaspectratio,viewport= 0 0 7in 6.32in,clip,scale=0.5]{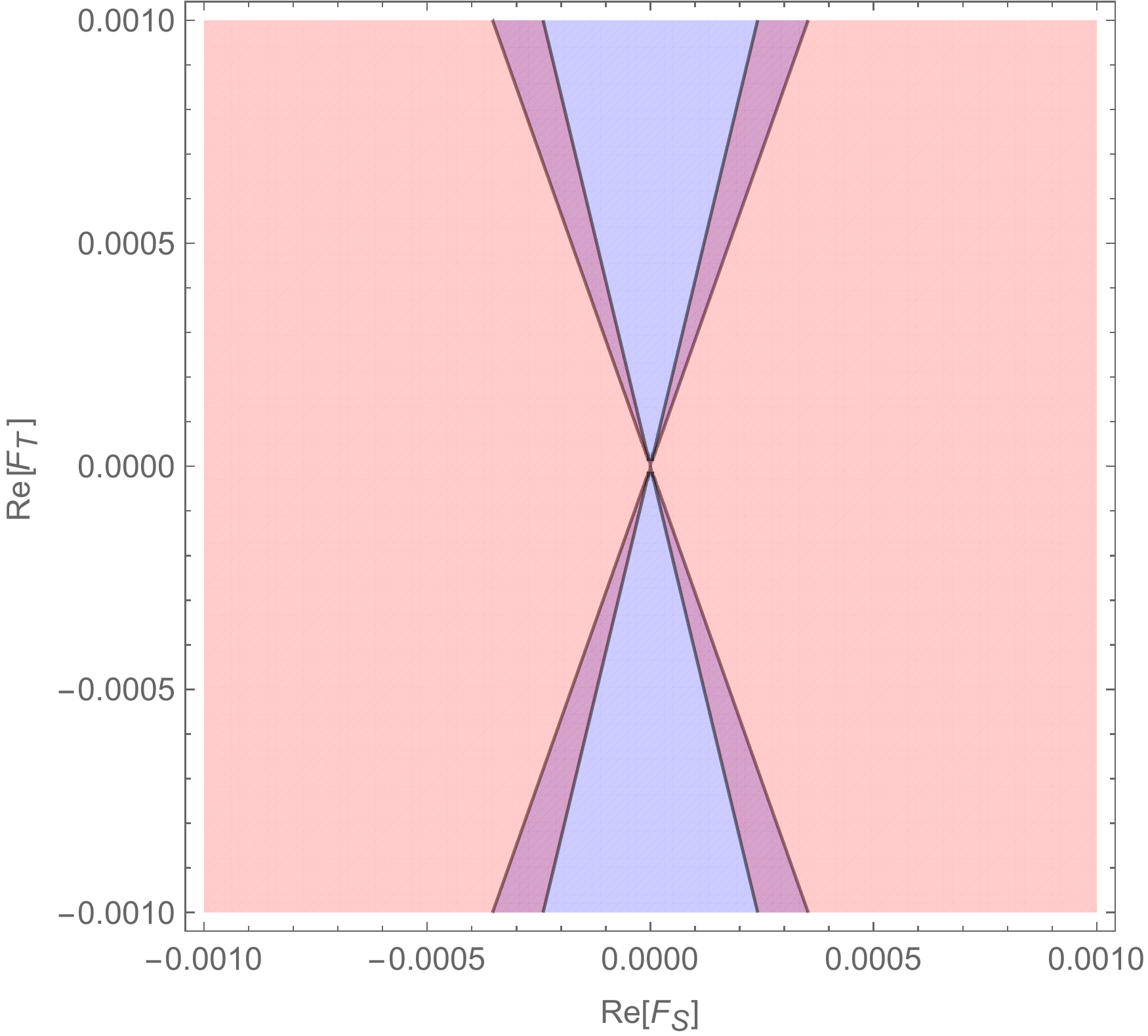}
  \caption{{\it Stable (red) and dS (blue) points for solutions close to a SUSY Mkw point (in the origin). The axes correspond to $\mathrm{Re}[F_{S}] \, = \, \epsilon$ and $\mathrm{Re}[F_{T}] \, = \, \lambda_T \, \epsilon$.}}
  \label{SUSYPlot}
\end{figure}

If $\lambda_T$ is left arbitrary, then these solutions become a set parameterised solely by $\epsilon$ and $\lambda_T$. To see their behaviour, we study the values of the cosmological constant and stability regions in a plot of $\mathrm{Re}[F_{T}]$ vs $\mathrm{Re}[F_{S}]$ (see figure~\ref{SUSYPlot}). This plot can be seen as a polar plot, in which the origin corresponds to a SUSY Mkw point. From (\ref{SUSYExFs}) we see that in the plot $\arctan (\lambda_T)$ is the corresponding polar angle and any point lays at a radial distance of $\sqrt{\mathrm{Re}[F_{S}] ^2 \, + \, \mathrm{Re}[F_{T}] ^2} = \sqrt{1 \, + \, \lambda_T^2 }\,\epsilon$ from a SUSY point.

The superposition of the stable regions (red) and dS regions (blue) is then clear and this overlap can be understood quite naturally as merely a manifestation of (\ref{SUSYbounds}). It should be pointed out that the origin of this plot represents a different SUSY Mkw solution for each direction in which one approaches it, i.e. the Mkw point depends on $\lambda_T$.

In order to be more concise, we present the explicit solution with $\lambda_T = 4$. The fluxes resulting fluxes are given in table~\ref{table:SUSYExFluxes}. Due to its linear dependence, the limit $\epsilon \rightarrow 0$ can be read quite easily. As it was discussed before, the eigenvalues of ${\left(m^{2}\right)^{I}}_{J}$ are organised by pairs. For $\lambda_T = 4$ they take the values

\be
\left\{\frac{1}{150} \, \epsilon^2 \,+ \, \mathcal{O}(\epsilon^{4})  \,\,\, , \,\,\, \frac{3511-31 \sqrt{6061}}{187272} \, + \, \mathcal{O}(\epsilon^{1}) \,\,\, , \,\,\, \frac{3511+31 \sqrt{6061}}{187272} \, + \, \mathcal{O}(\epsilon^{1}) \,\right\} \ ,
\ee

and the cosmological constant is then simply

\be
V \, = \, \frac{1}{12} \, \epsilon^2 \ .
\ee

\subsection*{Stable dS close to no-scale}
Two examples will be discussed of stable dS close to no-scale points. We will use a special kind of solutions connected to (\ref{Close2NSWs}), which were originally motivated by non-perturbative superpotentials of the kind of (\ref{W_nonpert}). Due to the simple mapping between 16 complex values of the $W$ derivatives and a 32 fluxes superpotential, it was possible to construct non-geometric polinomial superpotentials that share the same fundamental physical properties. We will provide explicit solutions making emphasis on this equivalence.

For these examples it is sufficient to consider superpotentials with the functional form of (\ref{W_nonpert})

\be \label{WPertExamps}
W_{\textrm{non-pert.}} = \,(P_{F} - P_{H} \, S )\,  + \, P_{Z} \, e^{i\,\a T} \ ,
\ee 

with

\be
P_Z \, = \, \left( \alpha_0 \, + \, i \, \alpha_1 \right) + \left( \alpha_2 \, + \, i \, \alpha_3 \right) \, U \ ,
\ee

with $\alpha_0$, $\alpha_1$, $\alpha_2$ and $\alpha_3$ being real parameters. 
As it was mentioned, conditions (\ref{NoScalePointConditions}) define a no-scale point. We may also start, as we did in the previous examples, by establishing a particular behaviour for the supersymmetry braking parameters. By doing so, it can readily be seen that the $\mathcal{O}(\epsilon^0)$ terms in $\mathrm{Re}[F_T]$ characterise the $\mathcal{O}(\epsilon^0)$ values of $W_0$ via $W_T = 0$ and, via equations of motion or (\ref{NoScalePointConditions}), the $\mathcal{O}(\epsilon^0)$ values of $W_S$ and $W_U$ as well. Hence, with this approach much of the freedom left is in the higher order terms in $\epsilon$.

In the present examples we will \emph{not} enforce $V_{\alpha \beta} = 0$. This is partly motivated by the fact that we want to show some explicit realizations of no-scale points with 3 massless modes. But most importantly, in the superpotential (\ref{WPertExamps}) we have 12 real fluxes which naturally fit with a description in terms of 6 real supersymmetry breaking parameters and 6 real constraints placed by the equations of motion (see \cite{Danielsson:2012by}). Hence, once picked $F_\alpha$ with the required $\epsilon \rightarrow 0$ limit, exact and well defined solutions can be found, with $\epsilon$ parameterizing a continuous trajectory from the chosen Mkw point. 

In this case it will be sufficient to break supersymmetry as
\be \label{NScaleExFs}
F_{\a} \, = \, \big( \, \epsilon  \, ,  \, 1 \, ,  \,  \lambda_U \, \epsilon \,  \big)\ .
\ee

Solutions corresponding to this choice are then parameterised in terms of $\epsilon$, $\lambda_U$ and the exponent factor in the non-perturbative term $\alpha$. The $\mathcal{O}(\epsilon^0)$ behaviour will be naturally of the no-scale type. This is all the background we need to find examples of stable dS close to no-scale points with 2 or 3 massless modes. In analogy with the SUSY case, $\lambda_U$ will help us parameterise the direction in which we approach no-scale points. In particular, it will determine whether we land in a 2 or 3 massless Mkw. In addition, these no-scale points will be of the GKP type. We will present simultaneously their corresponding manifestations as 32 fluxes models, in which, interestingly, only 16 will end up being different from 0. Hence, through this map, we will find for free non-geometric polynomial superpotentials with 16 fluxes that show dS close to a no-scale point.

For instance, if one picks $\alpha = \tfrac{2}{5}$, very interesting physics occurs. The corresponding solution in terms of the 12 fluxes of the non-perturbative superpotential (\ref{WPertExamps}) is given in the left part of table~\ref{table:NoSc2MExFluxesNONPERT} for $\lambda_U = 2$. As explained, we may compute the corresponding $W$ derivatives and then map them into the 32 fluxes non-geometric superpotential. The resulting solution is given in table~\ref{table:NoSc2MExFluxesNONGEO}. The fact that only 16 are non-zero is related to the fact that the $W$ derivatives have a very particular parity: $W_0$, $W_{\alpha \beta}$ are all imaginary while $W_\alpha$ and $W_{\alpha \beta \gamma}$ are all real (notice that the chosen supersymmetry braking parameters are all real).

\begin{figure}
  \centering
       \includegraphics[bb=0 0 8.50in 7.74in,keepaspectratio,viewport= 0 0 8.50in 7.74in,clip,scale=0.32]{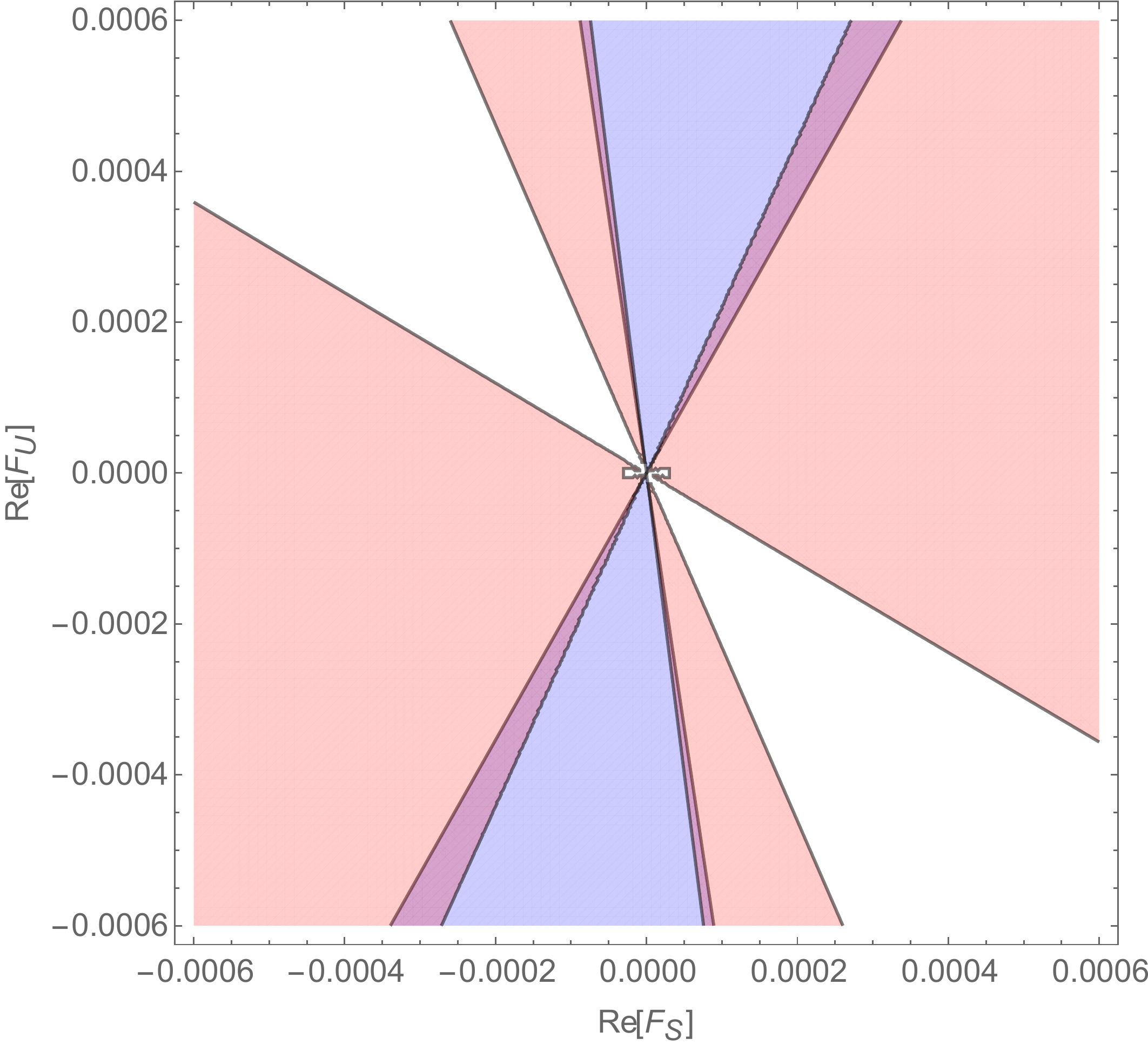}
      \ \ \ \ \ \ 
\includegraphics[bb=0 0 8.50in 8.31in,keepaspectratio,viewport= 0 0 8.50in 8.31in,clip,scale=0.3]{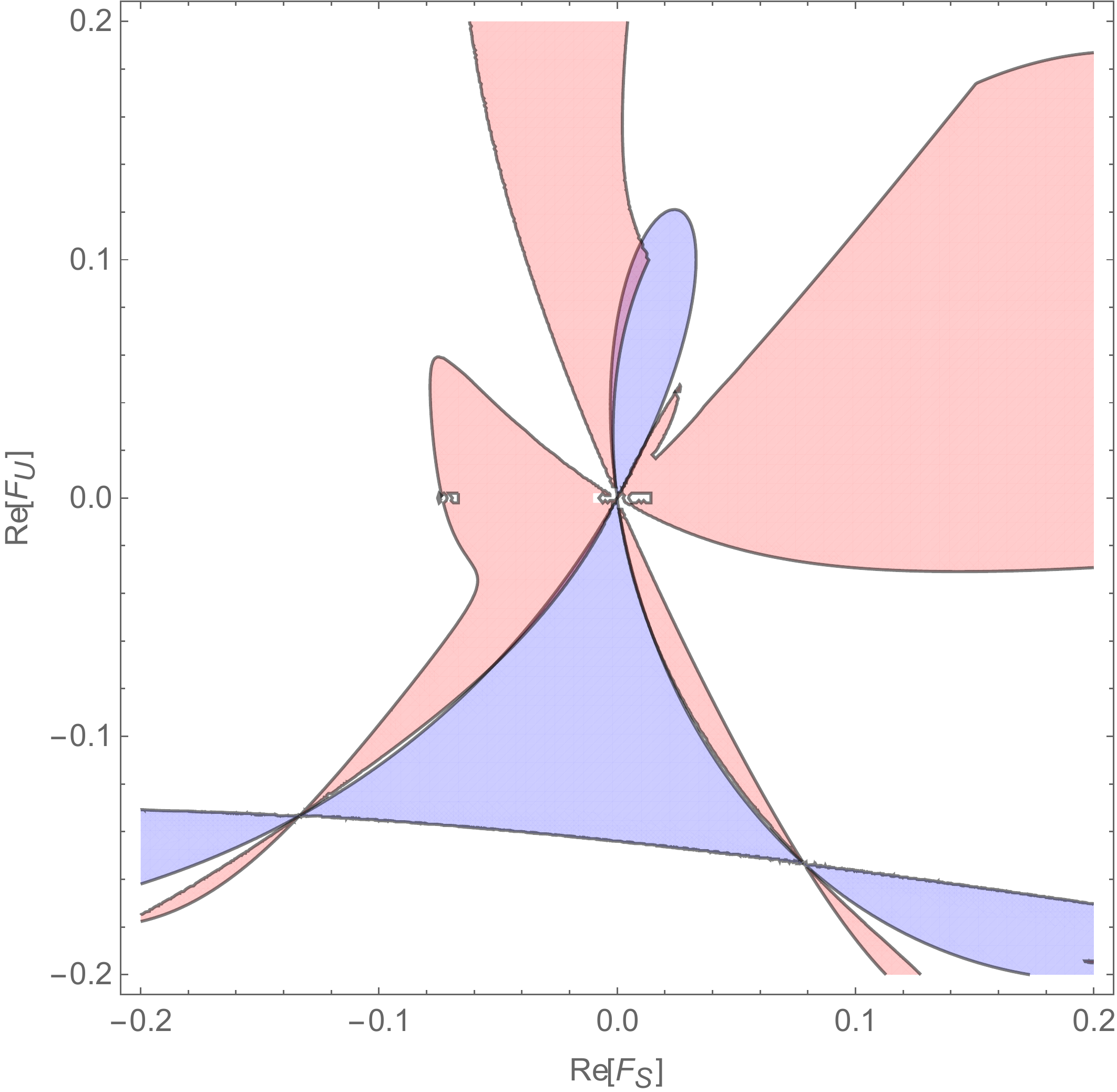}
  \caption{{\it Stable (red) and dS (blue) points for solutions close to a no-scale point (in the origin). The axes correspond to $\mathrm{Re}[F_{S}] \, = \, \epsilon$ and $\mathrm{Re}[F_{U}] \, = \, \lambda_U \, \epsilon$. Each plot shows a different scale. The regions of overlap are connected no-scale points with 2 massless modes.}}
  \label{NoScale2MPlots}
\end{figure}

\begin{table}
\renewcommand{\arraystretch}{1.25}
\begin{center}
\scalebox{0.92}[0.92]{
\begin{tabular}{ | c | c |}
\hline
labels &  values\\
\hline
\hline
$ a_0 $ & $ 0 $ \\
\hline
$ a_1 $ & $  \frac{38425 \epsilon ^3+7045 \epsilon ^2+990 \epsilon +24}{7800 \epsilon ^2+8880 \epsilon +576} $ \\
\hline
$ a_2 $ & $  0 $ \\
\hline
$ a_3 $ & $ \frac{375 \epsilon ^3+4215 \epsilon ^2-566 \epsilon -120}{24 \left(325 \epsilon ^2+370 \epsilon +24\right)} $ \\
 \hline
\hline
$ b_0 $ & $ \frac{-2175 \epsilon ^3-600 \epsilon ^2+877 \epsilon +60}{3900 \epsilon ^2+4440 \epsilon +288} $ \\
\hline
$ b_1 $ & $ 0 $ \\
\hline
$ b_2 $ & $ -\frac{1525 \epsilon ^3-740 \epsilon ^2+105 \epsilon +12}{3900 \epsilon ^2+4440 \epsilon +288} $ \\
\hline
$ b_3 $ & $  0 $ \\
\hline
\hline
$\alpha_0 $ & $ 0 $ \\
\hline
$ \alpha_1 $ & $ -\frac{25 e^{2/5} \epsilon  \left(105 \epsilon ^2+19 \epsilon -12\right)}{4 \left(325 \epsilon ^2+370 \epsilon +24\right)} $ \\
\hline
$ \alpha_2 $ & $ \frac{75 e^{2/5} \epsilon  \left(245 \epsilon ^2+17 \epsilon -4\right)}{4 \left(325 \epsilon ^2+370 \epsilon +24\right)} $ \\
\hline
$ \alpha_3 $ & $ 0 $ \\
 \hline
\end{tabular} \qquad
\begin{tabular}{ | c | c |}
\hline
labels & values \\
\hline
\hline
$ a_0 $ & $ 0 $ \\
\hline
$ a_1 $ & $ \frac{\epsilon  (1315 \epsilon +506)}{33 (20 \epsilon +31)} $ \\
\hline
$ a_2 $ & $  0 $ \\
\hline
$ a_3 $ & $ \frac{75 \epsilon ^2+25 \epsilon -31}{60 \epsilon +93} $ \\
 \hline
\hline
$ b_0 $ & $ \frac{-75 \epsilon ^2-25 \epsilon +31}{60 \epsilon +93} $ \\
\hline
$ b_1 $ & $ 0 $ \\
\hline
$ b_2 $ & $ \frac{(16-25 \epsilon ) \epsilon }{60 \epsilon +93} $ \\
\hline
$ b_3 $ & $  0 $ \\
\hline
\hline
$\alpha_0 $ & $ 0 $ \\
\hline
$ \alpha_1 $ & $ 0 $ \\
\hline
$ \alpha_2 $ & $ \frac{600 e^{11/10} \epsilon ^2}{220 \epsilon +341} $ \\
\hline
$ \alpha_3 $ & $ 0 $ \\
 \hline
\end{tabular}
}
\end{center}
\caption{{\it Two examples of stable dS solutions analytically connected to a 2-massless (left) and 3-massless (right) no-scale Mkw point realised within a non-perturbative superpotential of the type in \eqref{W_nonpert}.}}
\label{table:NoSc2MExFluxesNONPERT}
\end{table}

%\label{table:NoSc2MExFluxesNONPERT}
%\label{table:NoSc3MExFluxesNONPERT}

\begin{table}
\renewcommand{\arraystretch}{1.25}
\begin{center}
\scalebox{0.92}[0.92]{
\begin{tabular}{ | c | c |}
\hline
flux labels & flux values \\
\hline
\hline
$ a_0 $ & $ 0 $ \\
\hline
$ a_1 $ & $  \frac{-201065 \epsilon ^3+193449 \epsilon ^2+39660 \epsilon +960}{960 \left(325 \epsilon ^2+370 \epsilon +24\right)} $ \\
\hline
$ a_2 $ & $  0 $ \\
\hline
$ a_3 $ & $\frac{229515 \epsilon ^3+189541 \epsilon ^2-31460 \epsilon -4800}{960 \left(325 \epsilon ^2+370 \epsilon +24\right)} $ \\
 \hline
\hline
$ b_0 $ & $ \frac{-45165 \epsilon ^3-6811 \epsilon ^2+39740 \epsilon +4800}{960 \left(325 \epsilon ^2+370 \epsilon +24\right)} $ \\
\hline
$ b_1 $ & $ 0 $ \\
\hline
$ b_2 $ & $ -\frac{10385 \epsilon ^3-73001 \epsilon ^2+15540 \epsilon +960}{960 \left(325 \epsilon ^2+370 \epsilon +24\right)} $ \\
\hline
$ b_3 $ & $  0 $ \\
\hline
\hline
$ c_0 $ & $ -\frac{\epsilon  \left(114205 \epsilon ^2+14707 \epsilon -7820\right)}{320 \left(325 \epsilon ^2+370 \epsilon +24\right)} $ \\
\hline
$ c_1 $ & $ 0 $ \\
\hline
$ c_2 $ & $ -\frac{\epsilon  \left(65765 \epsilon ^2-2149 \epsilon +4820\right)}{320 \left(325 \epsilon ^2+370 \epsilon +24\right)} $ \\
\hline
$ c_3 $ & $ 0 $ \\
 \hline
\hline
$ d_0 $ & $ 0 $ \\
\hline
$ d_1 $ & $ -\frac{\epsilon  \left(31465 \epsilon ^2-4529 \epsilon +5380\right)}{320 \left(325 \epsilon ^2+370 \epsilon +24\right)} $ \\
\hline
$ d_2 $ & $ 0 $ \\
\hline
$ d_3 $ & $ -\frac{\epsilon  \left(111615 \epsilon ^2+13801 \epsilon -7140\right)}{960 \left(325 \epsilon ^2+370 \epsilon +24\right)}  $ \\
\hline
\end{tabular} \quad
\begin{tabular}{ | c | c |}
\hline
flux labels & flux values \\
\hline
\hline
$ a_0' $ & $ 0 $ \\
\hline
$ a_1' $ & $ -\frac{\epsilon  \left(8715 \epsilon ^2+6661 \epsilon -5460\right)}{960 \left(325 \epsilon ^2+370 \epsilon +24\right)} $ \\
\hline
$ a_2' $ & $ 0 $ \\
\hline
$ a_3' $ & $ -\frac{\epsilon  \left(191835 \epsilon ^2+18149 \epsilon -7380\right)}{960 \left(325 \epsilon ^2+370 \epsilon +24\right)} $ \\
\hline
\hline
$ b_0' $ & $ \frac{\epsilon  \left(128835 \epsilon ^2+41189 \epsilon -30420\right)}{960 \left(325 \epsilon ^2+370 \epsilon +24\right)} $ \\
\hline
$ b_1' $ & $ 0 $ \\
\hline
$ b_2' $ & $ \frac{\epsilon  \left(111615 \epsilon ^2+13801 \epsilon -7140\right)}{960 \left(325 \epsilon ^2+370 \epsilon +24\right)} $ \\
\hline
$ b_3' $ & $ 0 $ \\
\hline
\hline
$ c_0' $ & $ \frac{\epsilon  \left(945 \epsilon ^2+2303 \epsilon -1980\right)}{320 \left(325 \epsilon ^2+370 \epsilon +24\right)} $ \\
\hline
$ c_1' $ & $ 0 $ \\
\hline
$ c_2' $ & $ \frac{\epsilon  \left(20685 \epsilon ^2-4621 \epsilon +4980\right)}{320 \left(325 \epsilon ^2+370 \epsilon +24\right)} $ \\
\hline
$ c_3' $ & $ 0 $ \\
\hline
\hline
$ d_0' $ & $ 0 $ \\
\hline
$ d_1' $ & $ -\frac{\epsilon  \left(31465 \epsilon ^2-4529 \epsilon +5380\right)}{320 \left(325 \epsilon ^2+370 \epsilon +24\right)}$ \\
\hline
$ d_2' $ & $ 0 $ \\
\hline
$ d_3' $ & $ -\frac{\epsilon  \left(111615 \epsilon ^2+13801 \epsilon -7140\right)}{960 \left(325 \epsilon ^2+370 \epsilon +24\right)} $ \\
\hline
\end{tabular}
}
\end{center}
\caption{{\it An example of a non-geometric superpotential which analytically connects a 2-massless no-scale Mkw point with stable dS. This is the result of  mapping the first solution given in table~\ref{table:NoSc2MExFluxesNONPERT} into the 32 fluxes model.}}
\label{table:NoSc2MExFluxesNONGEO}
\end{table}

With $\epsilon$ and $\lambda_U$ arbitrary, we may make plots looking for stability and dS points around the no-scale point. In figure~\ref{NoScale2MPlots} we show $Re[F_S]$ vs $Re[F_U]$. As before, the origin corresponds to a Mkw point while $\arctan (\lambda_U)$ corresponds to the polar angle. The origin is a distinct no-scale point depending on the value of $\lambda_U$. In particular, if $\lambda_U$ takes any of the values $-3$, $-1$ , $1$ or $3$, the no-scale point has 3 massless modes. Any other direction lands in a no-scale with 2 massless modes.

For this value of $\alpha$, it can be seen that none of the 3-massless directions lands in stable dS (i.e. they land in stable Ads, unstable dS or unstable AdS). The superposition of the stable (red) and dS (blue) regions is clear and a particular realization of stable dS close to a no-scale with 2 massless modes is found when $\lambda_U = 2$. Notice also that, once we zoom out from the no-scale point, interesting phenomena are also manifest around other Mkw points. The cosmological constant with $\alpha = \tfrac{4}{10}$ and $\lambda_U = 2$ is

\bea
V &=& -\frac{\epsilon ^2 \left(870625 \epsilon ^4-203875 \epsilon ^3+78200 \epsilon ^2+5240 \epsilon -96\right)}{48 \left(325 \epsilon ^2+370 \epsilon +24\right)^2}
\\
&=& \frac{1}{288} \epsilon ^2 -\frac{1025}{3456} \, \epsilon ^3 \, + \,\mathcal{O}(\epsilon^4).
\eea

while the eigenvalues of ${\left(m^{2}\right)^{I}}_{J}$ take the values

\bea
&& \frac{79}{3240} \, \epsilon^2 \,+ \, \mathcal{O}(\epsilon^{3})  \,\,\, , \,\,\,
\frac{23}{2700} \, \epsilon^2 \,+ \, \mathcal{O}(\epsilon^{3})  \,\,\, , \,\,\, \frac{1}{1152} \, + \, \mathcal{O}(\epsilon^{1}) \,\,\, , \,\,\,
\\
&& \frac{1}{1152} \, + \, \mathcal{O}(\epsilon^{1}) \,\,\, , \,\,\,\,\,\,\,\,
\frac{25}{1152} \, + \, \mathcal{O}(\epsilon^{1}) \,\,\,\, \mathrm{and} \,\,\,\, \frac{1}{128} \, + \, \mathcal{O}(\epsilon^{1}) \, ,
\eea

The superpotential

\be
W \, = \frac{1}{24} \, \left(\,3 \, S U^2 \,- \,5 \, S\,+\,5 \, U^3\,-\,3 \, U\,\right)+ \mathcal{O}(\epsilon^{1}) \,,
\ee

is clearly of the GKP type at $\mathcal{O}(\epsilon^0)$.

\begin{table}
\renewcommand{\arraystretch}{1.25}
\begin{center}
\scalebox{0.92}[0.92]{
\begin{tabular}{ | c | c |}
\hline
flux labels & flux values \\
\hline
\hline
$ a_0 $ & $ 0 $ \\
\hline
$ a_1 $ & $  \frac{23 (160-31 \epsilon ) \epsilon }{240 (20 \epsilon +31)} $ \\
\hline
$ a_2 $ & $  0 $ \\
\hline
$ a_3 $ & $ \frac{8253 \epsilon ^2+2000 \epsilon -2480}{240 (20 \epsilon +31)} $ \\
 \hline
\hline
$ b_0 $ & $ \frac{-8253 \epsilon ^2-2000 \epsilon +2480}{4800 \epsilon +7440} $ \\
\hline
$ b_1 $ & $ 0 $ \\
\hline
$ b_2 $ & $ \frac{(1280-1007 \epsilon ) \epsilon }{240 (20 \epsilon +31)} $ \\
\hline
$ b_3 $ & $  0 $ \\
\hline
\hline
$ c_0 $ & $ -\frac{751 \epsilon ^2}{80 (20 \epsilon +31)} $ \\
\hline
$ c_1 $ & $ 0 $ \\
\hline
$ c_2 $ & $ -\frac{1833 \epsilon ^2}{80 (20 \epsilon +31)} $ \\
\hline
$ c_3 $ & $ 0 $ \\
 \hline
\hline
$ d_0 $ & $ 0 $ \\
\hline
$ d_1 $ & $  -\frac{1413 \epsilon ^2}{80 (20 \epsilon +31)} $ \\
\hline
$ d_2 $ & $ 0 $ \\
\hline
$ d_3 $ & $ -\frac{331 \epsilon ^2}{80 (20 \epsilon +31)} $ \\
\hline
\end{tabular} \quad
\begin{tabular}{ | c | c |}
\hline
flux labels & flux values \\
\hline
\hline
$ a_0' $ & $ 0 $ \\
\hline
$ a_1' $ & $ \frac{89 \epsilon ^2}{80 (20 \epsilon +31)} $ \\
\hline
$ a_2' $ & $ 0 $ \\
\hline
$ a_3' $ & $ -\frac{509 \epsilon ^2}{80 (20 \epsilon +31)} $ \\
\hline
\hline
$ b_0' $ & $ -\frac{751 \epsilon ^2}{80 (20 \epsilon +31)} $ \\
\hline
$ b_1' $ & $ 0 $ \\
\hline
$ b_2' $ & $ \frac{331 \epsilon ^2}{80 (20 \epsilon +31)} $ \\
\hline
$ b_3' $ & $ 0 $ \\
\hline
\hline
$ c_0' $ & $ -\frac{331 \epsilon ^2}{80 (20 \epsilon +31)} $ \\
\hline
$ c_1' $ & $ 0 $ \\
\hline
$ c_2' $ & $ \frac{227 \epsilon ^2}{80 (20 \epsilon +31)} $ \\
\hline
$ c_3' $ & $ 0 $ \\
\hline
\hline
$ d_0' $ & $ 0 $ \\
\hline
$ d_1' $ & $ -\frac{1413 \epsilon ^2}{80 (20 \epsilon +31)} $ \\
\hline
$ d_2' $ & $ 0 $ \\
\hline
$ d_3' $ & $ -\frac{331 \epsilon ^2}{80 (20 \epsilon +31)} $ \\
\hline
\end{tabular}
}
\end{center}
\caption{{\it An example of a non-geometric superpotential which analytically connects a 3-massless no-scale Mkw point with stable dS. This is the result of mapping the second solution given in table~\ref{table:NoSc2MExFluxesNONPERT} into the 32 fluxes model.}}
\label{table:NoSc3MExFluxesNONGEO}
\end{table}

\begin{figure}
  \centering
       \includegraphics[bb=0 0 11.97in 5.32in,keepaspectratio,viewport= 1.2in 0 11.97in 5.32in,clip,scale=0.33]{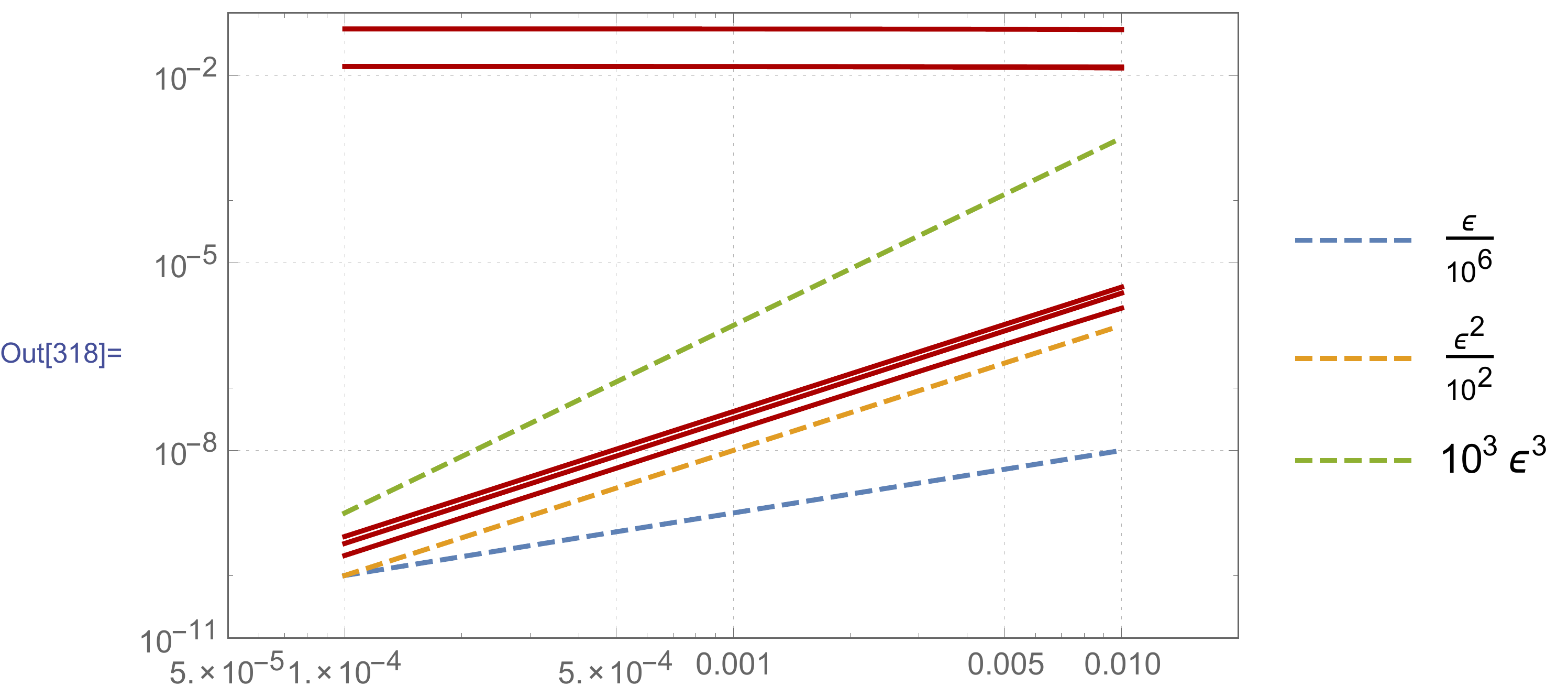}
  \caption{{\it Eigenvalues of ${\left(m^{2}\right)^{I}}_{J}$ (red) as a function of $\epsilon$, shown in a loglog plot for a stable dS solution close to a no-scale point with 3 massless states. For reference, linear (blue), quadratic (yellow) and cubic (green) dependences are shown. Two of the massive states share a similar value (around $10^{-2}$).}}
  \label{NoScale3MPlot}
\end{figure}

To see an example of stable dS close to a no-scale wth 3-massless modes one can simple take $\alpha = \tfrac{11}{10}$ and $\lambda_U = 1$. The solutions in terms of the 12 parameters in the non-perturbative superpotential can be seen in the right part of table~\ref{table:NoSc2MExFluxesNONPERT} while its non-geometric counterpart is given in table~\ref{table:NoSc3MExFluxesNONGEO}. It shares many of the features we discussed previously. The corresponding potential is then

\bea
V &=&\frac{\epsilon ^2 \left(-500 \epsilon ^2+640 \epsilon +31\right)}{24 (20 \epsilon +31)^2}
\\
&=& \frac{1}{744} \epsilon ^2 +\frac{25}{961} \, \epsilon ^3 \, + \,\mathcal{O}(\epsilon^4).
\eea
exhibits, as expected, a simple GKP form.

and the scaling behaviour of the eigenvalues of ${\left(m^{2}\right)^{I}}_{J}$  can be observed in figure~\ref{NoScale3MPlot}. Notice that in this case all the massless modes grow quadratically. We will discuss an example with a mass growing linearly with $\epsilon$ in the next subsection. The  $\mathcal{O}(\epsilon^0)$ superpotential

\be
W \, = \frac{1}{3}\, \left(\,U^3\,-\,S\,\right) + \mathcal{O}(\epsilon^{1}) \ ,
\ee
exhibits, as expected, a simple GKP form.

It is worth mentioning that there exist plenty of approximately no-scale stable dS solutions which can be constructed analytically without the aid of explicit non-perturbative terms by
using the procedure sketched in section~\ref{sec:method} which may be combined with solving the pairwise degeneracy constraint for the mass matrix $V_{\a\b}\,=\,0$. These solutions generically approach no-scale Mkw points with two flat directions, but of a generalised type (see appendix~\ref{app:no-scale_Mkw}).   

\subsection*{Massless Mode lifted at linear order}
This particular example shows a solution that moves continuously from a no-scale Mkw point with 3 massless modes to an unstable dS. One of the massless modes is lifted at $\mathcal{O}(\epsilon^1)$ and becomes positive for positive $\epsilon$. Nevertheless, there is a massless mode that becomes tachyonic with a negative $\mathcal{O}(\epsilon^2)$ term. This specific solution is given in table~\ref{table:LinearMass}.
\begin{table}
\renewcommand{\arraystretch}{1.25}
\begin{center}
\scalebox{0.92}[0.92]{
\begin{tabular}{ | c | c |}
\hline
flux labels & flux values \\
\hline
\hline
$ a_0 $ & $ \frac{21 \epsilon ^3}{8}-\frac{21 \epsilon ^2}{4}+\frac{17 \epsilon }{8}-\frac{13}{48} $ \\
\hline
$ a_1 $ & $ \frac{7 \epsilon ^3}{4}-\frac{11 \epsilon ^2}{6}+\frac{3 \epsilon }{8}+\frac{1}{8} $ \\
\hline
$ a_2 $ & $ -\frac{7 \epsilon ^3}{8}-\frac{\epsilon ^2}{4}+\frac{3 \epsilon }{8}-\frac{7}{48} $ \\
\hline
$ a_3 $ & $ -\epsilon ^2+\frac{3 \epsilon }{8}-\frac{1}{24} $ \\
\hline
\hline
$ b_0 $ & $ \frac{7 \epsilon ^3}{4}-\frac{9 \epsilon ^2}{4}+\frac{\epsilon }{8}+\frac{1}{8} $ \\
\hline
$ b_1 $ & $ -\frac{7 \epsilon ^3}{8}+\frac{3 \epsilon }{8}-\frac{3}{16} $ \\
\hline
$ b_2 $ & $ -\frac{11 \epsilon ^2}{12}+\frac{7 \epsilon }{24}-\frac{1}{24} $ \\
\hline
$ b_3 $ & $ -\frac{7 \epsilon ^3}{8}+\frac{\epsilon ^2}{2}+\frac{\epsilon }{8}-\frac{1}{16} $ \\
\hline
\hline
$ c_0 $ & $ \frac{7 \epsilon ^3}{6}-\frac{\epsilon ^2}{4}-\frac{17 \epsilon }{24}+\frac{5}{24} $ \\
\hline
$ c_1 $ & $ \frac{21 \epsilon ^3}{8}-\frac{11 \epsilon ^2}{6}+\frac{\epsilon }{12}-\frac{3}{16} $ \\
\hline
$ c_2 $ & $ \frac{7 \epsilon ^3}{4}-\frac{19 \epsilon ^2}{12}+\frac{11 \epsilon }{24}-\frac{1}{8} $ \\
\hline
$ c_3 $ & $ \frac{7 \epsilon ^3}{24}-\frac{\epsilon }{12}+\frac{1}{48} $ \\
\hline
\hline
$ d_0 $ & $ -\frac{7 \epsilon ^3}{8}+\frac{7 \epsilon ^2}{12}-\frac{\epsilon }{4}+\frac{1}{48} $ \\
\hline
$ d_1 $ & $ \frac{7 \epsilon ^3}{4}-\frac{5 \epsilon ^2}{3}+\frac{13 \epsilon }{24}-\frac{1}{8} $ \\
\hline
$ d_2 $ & $ -\frac{7 \epsilon ^3}{8}+\frac{\epsilon ^2}{4}+\frac{\epsilon }{4}-\frac{3}{16} $\\
\hline
$ d_3 $ & $ \frac{5 \epsilon ^2}{6}-\frac{7 \epsilon }{24}+\frac{1}{24} $ \\
\hline
\end{tabular} \quad
\begin{tabular}{ | c | c |}
\hline
flux labels & flux values \\
\hline
\hline
$ a_0' $ & $ \frac{7 \epsilon ^3}{8}-\epsilon ^2+\frac{1}{48} $ \\
\hline
$ a_1' $ & $ \frac{7 \epsilon ^2}{12}-\frac{5 \epsilon }{24}+\frac{1}{24} $ \\
\hline
$ a_2' $ & $ \frac{7 \epsilon ^3}{8}-\frac{5 \epsilon ^2}{6}+\frac{1}{16} $ \\
\hline
$ a_3' $ & $ \frac{7 \epsilon ^3}{4}+\frac{\epsilon ^2}{4}-\frac{3 \epsilon }{8}+\frac{1}{24} $ \\
\hline
\hline
$ b_0' $ & $ \frac{7 \epsilon ^3}{4}-\frac{7 \epsilon ^2}{2}+\frac{9 \epsilon }{8}-\frac{5}{24} $ \\
\hline
$ b_1' $ & $ -\frac{7 \epsilon ^3}{8}+\frac{3 \epsilon ^2}{4}-\frac{1}{16} $ \\
\hline
$ b_2' $ & $ -\frac{2 \epsilon ^2}{3}+\frac{7 \epsilon }{24}-\frac{1}{24} $ \\
\hline
$ b_3' $ & $ -\frac{7 \epsilon ^3}{8}+\frac{3 \epsilon ^2}{4}-\frac{1}{48} $ \\
\hline
\hline
$ c_0' $ & $ \frac{7 \epsilon ^3}{12}-\frac{2 \epsilon ^2}{3}+\frac{5 \epsilon }{24}-\frac{1}{24} $ \\
\hline
$ c_1' $ & $ \frac{7 \epsilon ^3}{8}-\frac{5 \epsilon ^2}{4}+\frac{\epsilon }{8}-\frac{1}{16} $ \\
\hline
$ c_2' $ & $ \frac{5 \epsilon ^2}{6}-\frac{11 \epsilon }{24}+\frac{1}{8} $ \\
\hline
$ c_3' $ & $ -\frac{7 \epsilon ^3}{24}+\frac{17 \epsilon ^2}{12}-\frac{\epsilon }{8}-\frac{5}{48} $ \\
\hline
\hline
$ d_0' $ & $ -\frac{7 \epsilon ^3}{8}+\frac{\epsilon ^2}{2}+\frac{\epsilon }{8}-\frac{5}{48} $ \\
\hline
$ d_1' $ & $ \frac{7 \epsilon ^3}{4}-\frac{23 \epsilon ^2}{12}+\frac{13 \epsilon }{24}-\frac{1}{8} $ \\
\hline
$ d_2' $ & $ -\frac{7 \epsilon ^3}{8}+\epsilon ^2-\frac{\epsilon }{8}-\frac{1}{16} $ \\
\hline
$ d_3' $ & $ \frac{5 \epsilon ^2}{12}-\frac{7 \epsilon }{24}+\frac{1}{24} $ \\
\hline
\end{tabular}
}
\end{center}
\caption{{\it An example of a no-scale point with a massless mode growing linearly as it moves towards unstable dS.}}
\label{table:LinearMass}
\end{table}

\begin{figure}
  \centering
       \includegraphics[bb=0 0 11.6in 5.39in,keepaspectratio,viewport=1.0in 0 11.6in 5.39in,clip,scale=0.3]{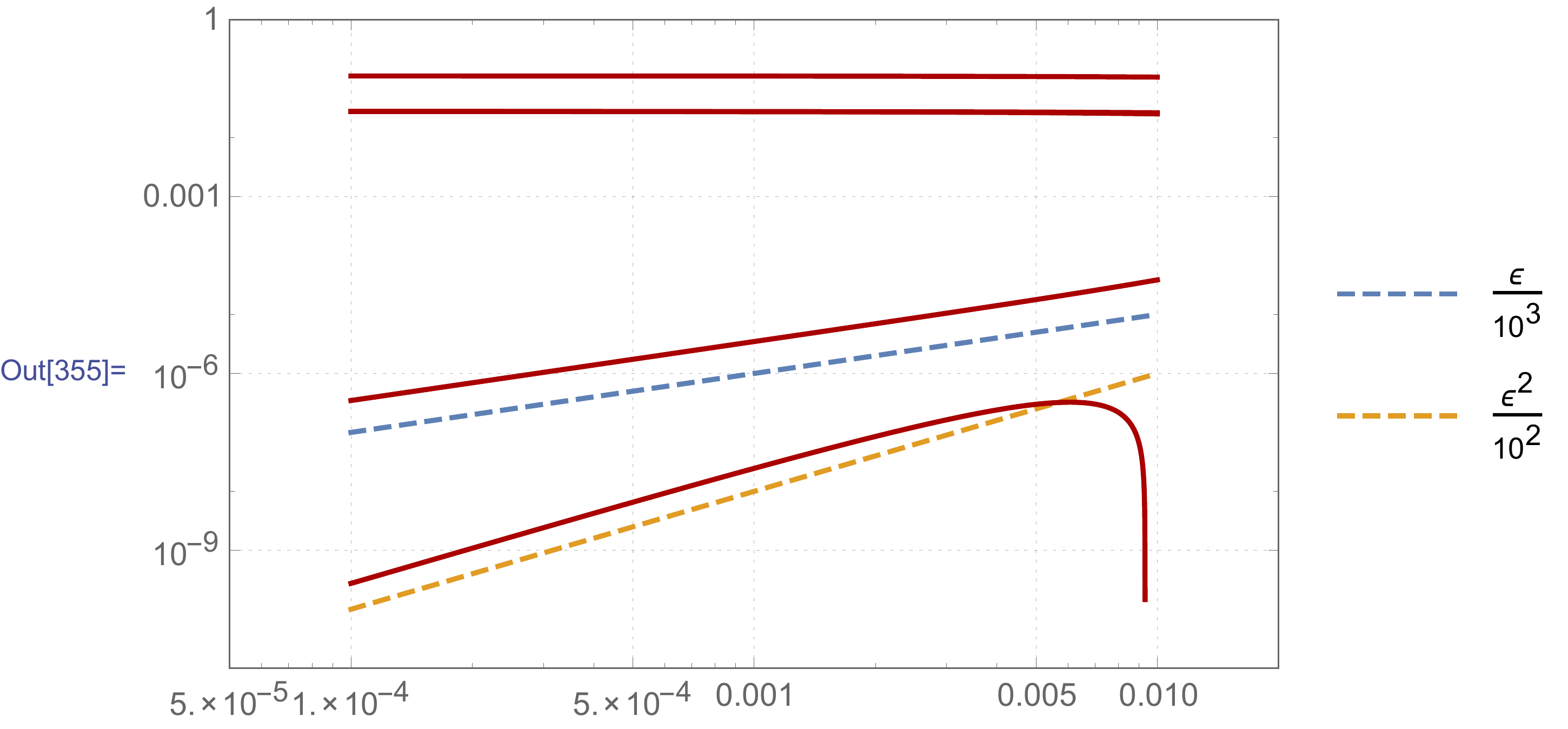}
      \ \ \ 
\includegraphics[bb=0 0 11.6in 5.32in,keepaspectratio,viewport= 1.2in 0 11.6in 5.32in,clip,scale=0.3]{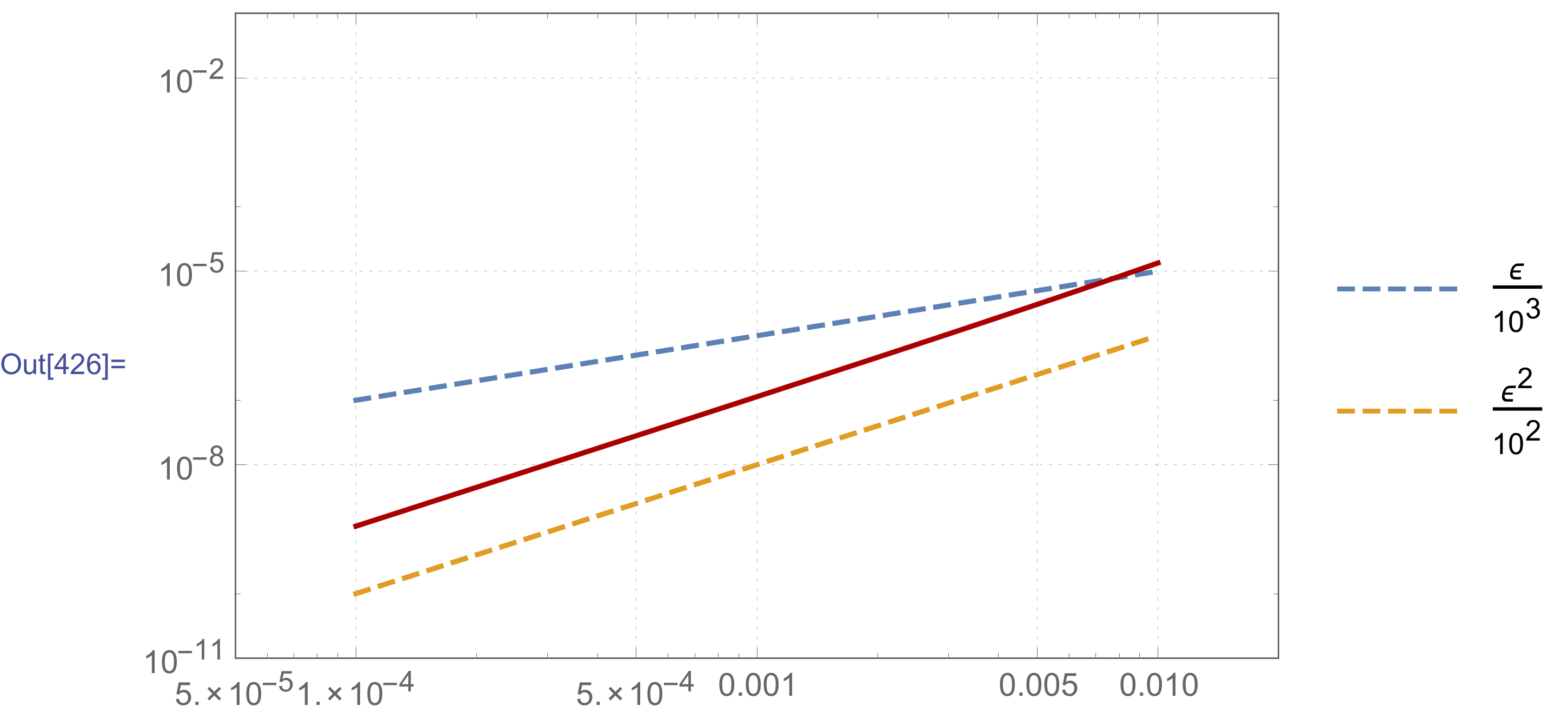}
  \caption{{\it Eigenvalues of ${\left(m^{2}\right)^{I}}_{J}$ (red) as a function of $\epsilon$, shown in a loglog plot for an unstable dS solution close to a no-scale point with 3 massless states. In the plot on the left we see the  5 positive masses (notice that one is linear). Two of the massive states share a similar value (around 0.003). The sixth mass is negative and in the plot on the right we plot its absolute value. For reference, linear (blue), and quadratic (yellow) dependences are shown.}}
  \label{LinearMass}
\end{figure}

While the cosmological constant takes the simple form
\be
V \, = \, \frac{\epsilon^2 }{12} \ ,
\ee
the described scaling behaviour of the masses can be observed in figure~\ref{LinearMass}.

\section{Conclusions}
\label{sec:conclusions}

In this paper we have shown how dS vacua can be constructed close to no-scale as well as supersymmetric Mkw points using a polynomial superpotential with non-geometric fluxes. As far as the existence of critical points and the features of the corresponding mass spectra are concerned, the  superpotential with generalised fluxes covers all possibilities, and explicit non-perturbative terms will not bring in anything new.

The case of supersymmetric Mkw is particularly interesting. In contrast to previous constructions, such as \cite{Kallosh:2014oja}, we did not need to add a Polonyi field to be able to break supersymmetry. Instead, we managed to arrange for flat directions already within the STU-model itself. However, whether it is actually possible to tune parameters to obtain the desired hierarchy with $V \ll m^2_{\rm gravitino} \ll m^2_{\rm scalars}$ still remains to be seen. The large mass of the Polonyi field of \cite{Kallosh:2014oja} was arranged through tuning of its K{\"a}hler potential. Since we have a large number of fluxes available we expect to be able to perform similar tunings in our case. It would be interesting to try to construct a phenomenologically viable example and we hope to come back to this issue in the future. 

In contrast to the non-perturbative terms the generalised fluxes are fully determined by string dualities and the number of parameters limited to $32$. Furthermore, a picture is emerging where at least some combinations of the non-geometric fluxes can be given a fully geometric interpretation through compactifications of string/M-theory on other topologies than twisted tori \cite{Danielsson:2015tsa}. Examples includes M-theory on $S^7$ and on $S^4 \times T^3$. 

It will be interesting to see whether any of the stable dS vacua that can be constructed using the techniques of this paper can be given a fully geometric interpretation. Which features of the topology will be crucial for success? The advantage of such a solution is that it will be under firm controll from within supergravity and not dependent upon non-perturbative and string-inspired corrections under limited control.

%%%%%%%%%%%%%%%%%%%%%%%%%%%%%%%%%%%%
%
% Acknowledgments
%
%%%%%%%%%%%%%%%%%%%%%%%%%%%%%%%%%%%%

\section*{Acknowledgments}

We would like to thank D.~Marsh, B.~Vercnocke and T.~Wrase for stimulating and valuable discussions. JB is supported by the John Templeton Foundation Grant 48222 and the CEA Eurotalents program. The work of UD, GD and SV is supported by the Swedish Research Council (VR).

\newpage

%%%%%%%%%%%%%%%%%%%%%%%%%%%%%%%%%%%%
%
% Appendices
%
%%%%%%%%%%%%%%%%%%%%%%%%%%%%%%%%%%%%

\appendix

\section{Classification of SUSY Mkw solutions}
\label{app:SUSY_Mkw}

We start from the duality-covariant superpotential in \eqref{W_all_fluxes} and study the most general SUSY Mkw solution within this STU-model. To this end, one needs to impose the following conditions 
\be
\begin{array}{lcclc}
W \ \overset{!}{=} \ 0 & , & & F_{S} \ = \ F_{T} \ = \ F_{U} \ \overset{!}{=} \ 0 & ,
\end{array}
\ee
Solutions to the above equations may be exhaustively studied in the origin of moduli space \eqref{origin}, where the problem reduces to a system of eight real linear homogenous equations in the $32$ 
superpotential couplings of tables~\ref{table:unprimed_fluxes} and \ref{table:primed_fluxes}. The complete space of solutions is then spanned by $24$ independent generalised fluxes, which can be 
used to construct the following explicit parametrisation of \emph{all} SUSY Mkw solutions
\be
\begin{array}{lclc}
a_{0} & = & -3 a_{2}' \, + \, b_3 \, + \, b_{3}' \, + \, 2 c_1 \, - \, c_1' \, - \, c_3 \, - \, 4 c_3' \, - \, 2 d_0 \, - \, 2 d_0' \, + \, d_2 \, + \, d_2' & , \\
a_0' & = & 3 a_2'  \, + \, 3 b_1' \, - \, b_3' \, - \, c_1  \, + \, 2 c_1' \, - \, c_3  \, + \, 2 c_3'  \, + \, d_0 - 2 d_0'  \, + \, d_2 \, - \, 2 d_2' & , \\
a_2 & = & -a_2' \, - \, b_1' \, + \, \frac{1}{3}b_3 \, + \, \frac{2}{3}b_3' \, - \, c_3 \, - \, c_3' \, - \, d_0 \, + \, d_2' & , \\
b_1 & = & b_1' \, + \, \frac{1}{3}b_3 \, - \, \frac{1}{3}b_3' \, + \, d_0 \, - \, d_0' \, + \, d_2 \, - \, d_2' & , \\
a_1 & = & -\frac{1}{3}a_3' \, - \, \frac{2}{3}b_0' \, - \, b_2 \, + \, b_2' \, + \, \frac{4}{3}c_0 \, + \, \frac{1}{3}c_0' \, + \, \frac{1}{3}c_2 \, - \, \frac{2}{3}c_2' \, - \, \frac{2}{3}d_1 \, + \, \frac{4}{3}d_1' \, - \, \frac{5}{3}d_3 \, + \,\frac{1}{3}d_3' & , \\
a_1' & = & \frac{1}{3}a_3' \, - \, \frac{1}{3}b_0' \, + \, b_2' \, - \, \frac{1}{3}c_0 \, + \, \frac{2}{3}c_0' \, - \,  \frac{1}{3}c_2 \, + \, \frac{2}{3}c_2' \, - \, \frac{1}{3}d_1 \, + \, \frac{2}{3}d_1' \, - \, \frac{1}{3}d_3 \, + \, \frac{2}{3}d_3' & , \\
a_3 & = & -a_3' \, - \, b_0' \, - \, 3 b_2 \, + \, 2 c_0 \, + \, 2 c_0' \, - \, c_2 \, - \, c_2' \, - \, d_1 \, + \, 2 d_1' \, - \, 4 d_3 \, - \, d_3' & , \\
b_0 & = & b_0' \, + \, 3 b_2 \, - \, 3 b_2' \, + \, 3 d_1 \, - \, 3 d_1' \, + \, 3 d_3 \, - \, 3 d_3' & .
\end{array}
\ee
Please note that the above SUSY Mkw solutions generically (\emph{i.e.} for arbitrarily chosen values of the $24$ parameters) have all six real scalars stabilised with a positive mass.
Due to the theorem in ref.~\cite{Kallosh:2014oja}, such non-degenerate SUSY Mkw points are isolated in the sense that they cannot be continuously deformed by breaking SUSY into \emph{e.g.} dS solutions.
This automatically implies that the SUSY Mkw that our stable dS can approach in the small-$\epsilon$ limit will always by construction be of the degenerate type, \emph{i.e.} with two flat directions.
Such directions acquire a mass when deforming the SUSY Mkw point into dS and hence exactly coincide with the sGoldstini directions.

\section{Classification of no-scale Mkw solutions}
\label{app:no-scale_Mkw}

Traditional no-scale supergravity models are naturally constructed within type IIB compactifications with both NS-NS and R-R gauge fluxes and yield a superpotential of the form \eqref{W_GKP}, which has
the special feature of being completely independent of the K\"ahler modulus $T$. Such GKP-like backgrounds produce stable Mkw solutions up to the two real directions sitting in $T$ which remain totally 
flat. Moreover, upon imposing a special degeneracy condition on the fluxes, it is actually possible to obtain exceptional no-scale models with an extra massless mode. 
In every no-scale model, SUSY is naturally broken by large amounts in the $T$ direction, whereas $F_{S}$ \& $F_{U}$ stay zero. 

In this appendix we present a class of STU-superpotentials that generalises that in \eqref{W_GKP} and still retains all the above defining features of a no-scale, except they have some $T$-dependence,
contrary to \eqref{W_GKP}. We will refer to these models as \emph{generalised no-scale models}. The most general Mkw solutions of the no-scale type within these models can be parametrised in terms
of 10 arbitrary real fluxes as
%cc
\be
\begin{array}{cc}
\begin{array}{lclc}
 a_0'  & = & c_1 -2 d_0'  & , \\
 a_1'  & = & -\frac{1}{3} d_1  & , \\
 a_2  & = & b_1 +2 d_0' -\frac{2}{3} c_1 -\frac{2}{3} d_2  & , \\
 a_2'  & = & -\frac{1}{3}d_2  & , \\
 a_3'  & = & -\frac{1}{2} b_0' + d_1'-\frac{1}{2} d_1  & , \\
 b_0  & = & - a_3 - b_0' + d_1  & , \\
 b_1'  & = & \frac{1}{3}c_1  & , \\
 b_2  & = & - a_1 -\frac{2}{3} d_1' & , \\
 b_2'  & = & \frac{1}{2} b_0' -\frac{1}{3}d_1' -\frac{1}{6} d_1  & , \\
 b_3  & = & a_0 -2 d_0' & , \\
 b_3'  & = & 2 d_0' - d_2   & , \\
 \end{array} \,\,\, & \,\,\,
\begin{array}{lclc}
c_0  & = & - b_0' & , \\
 c_0'  & = & \frac{1}{2} b_0' -\frac{1}{2} d_1  & , \\
 c_1'  & = & c_1 -3 d_0' + d_2 & , \\
 c_2  & = & \frac{3}{2} b_0' - d_1' - \frac{1}{2} d_1 & , \\
 c_2'  & = & - d_1'  & , \\
 c_3  & = & d_2 -2 d_0' & , \\
 c_3'  & = & d_0'  & , \\
 d_0  & = & c_1 -2 d_0'  & , \\
 d_2'  & = & c_1 -3 d_0' + d_2  & , \\
 d_3  & = & -\frac{1}{2} b_0' + d_1'-\frac{1}{2} d_1 & , \\
 d_3'  & = & \frac{1}{2} d_1 -\frac{1}{2} b_0' & . \\
\end{array}
\end{array}
\ee
The above Mkw solutions generically have two massless directions, but by imposing some extra finetuning condition, one can recover the degenerate case of a special no-scale Mkw solution with three
flat directions. By deforming these Mkw points (\emph{i.e.} by turning on a small $\epsilon$), one can construct dS solutions which lie arbitrarily close to them, both in the case of a generic and in the
case of a degenerate no-scale point and in either case the aforementioned flat directions get lifted by $\epsilon$-corrections.

%%%%%%%%%%%%%%%%%%%%%%%%%%%%%%%%%%%%
%
% Bibliography
%
%%%%%%%%%%%%%%%%%%%%%%%%%%%%%%%%%%%%

\small

\bibliography{references}
\bibliographystyle{utphys}

\end{document}